\documentclass[12pt]{article}

\title{Generalized orthogonal polynomials, discrete KP and
Riemann-Hilbert problems\footnote{Appeared in: Comm.
Math. Phys., 207, 589--620 (1999) }}

\author{M. Adler\thanks{Department of Mathematics,
Brandeis University, Waltham, Mass 02454, USA. E-mail:
adler@math.brandeis.edu. The support of a National Science
Foundation grant \# DMS-98-4-50790 is gratefully
acknowledged.}~~~~~~P. van Moerbeke\thanks{Department of
Mathematics, Universit\'e de Louvain, 1348 Louvain-la-Neuve,
Belgium and Brandeis University, Waltham, Mass 02454, USA. E-mail:
vanmoerbeke@geom.ucl.ac.be and @math.brandeis.edu. The  support of
a National Science Foundation grant \# DMS-98-4-50790, a Nato, a
FNRS and a Francqui Foundation grant is gratefully acknowledged.
Some of the present work was done at the Centre Emile Borel, Paris
(fall 96).}}

\date{August 24, 1998}

\newcommand{\MAT}[1]{\left(\begin{array}{*#1c}}
\newcommand{\mat}{\end{array}\right)}
\newcommand{\qed}
{%
\mbox{}%
\nolinebreak%
\hfill%
\rule{2mm}{2mm}%
\medbreak%
\par%
}

\newcommand{\sumbis}[2]%
{%

\begin{array}[t]{c}
\sum\\
{\scriptstyle #1}\\
{\scriptstyle #2}
\end{array}

}

\newcommand{\lrg}{\longrightarrow}

\newcommand{\BC}{{\Bbb C}}

\newcommand{\BZ}{{\Bbb Z}}

\newcommand{\iy}{\infty}
\newcommand{\pl}{\partial}
\newcommand{\al}{\alpha}
\newcommand{\proof}{\underline{\sl Proof}: }
\newcommand{\remark}{\underline{\sl Remark}: }

\newcommand{\HR}{{\cal H}}

\newcommand{\WR}{{\cal W}}

\newcommand{\la}{\langle}
\newcommand{\ra}{\rangle}

\newcommand{\dt}{\delta}
\newcommand{\Dt}{\Delta}

\newcommand{\sg}{\sigma}
\newcommand{\BR}{{\Bbb R}}
\newcommand{\lb}{\lambda}
\newcommand{\Lb}{\Lambda}
\newcommand{\tr}{\mbox{tr}}

\newcommand{\Span}{\mathop{\rm span}}
\newcommand{\diag}{\mathop{\rm diag}}

\def\be#1\ee{{\begin{equation} #1 \end{equation}}}
\def\bea#1\eea{{\begin{eqnarray} #1 \end{eqnarray}}}
\ifx\undefined\Bbb
        \let\Bbb\bf
\fi

\catcode`\@=11
\def\ps@X{\let\@mkboth\@gobbletwo
        \def\@oddhead{\tt Adler-van Moerbeke:%
        Polyn/RH-problems\hfil Aug 24, 1998\ \ \hfil\S\thesection,
p.\thepage
        }
        \def\@oddfoot{\rm\hfil\thepage\hfil}
        \let\@evenhead\@oddhead
        \let\@evenfoot\@oddfoot}
\catcode`@=12
\newtheorem{definition}{Definition}[
section]

\newtheorem{theorem}[definition]{Theorem}

\newtheorem{lemma}[definition]{Lemma}
\newtheorem{corollary}[definition]{Corollary}
\newtheorem{proposition}[definition]{Proposition}

\catcode`\@=11
\let\c@equation=\relax
\newcounter{equation}[section]

\catcode`\@=12

\begin{document}
\maketitle

\setcounter{section}{-1}
{\bf Dedicated to J\"urgen Moser, at the occasion of his 70th
birthday}

\begin{abstract}Classically, a single weight on an interval
 of the real line leads to moments, orthogonal
 polynomials and tridiagonal matrices.
 Appropriately deforming this weight with times
 $t=(t_1,t_2,...)$, leads to the standard Toda lattice
 and $\tau$-functions, expressed as Hermitian matrix integrals.

This paper is concerned with a sequence of $t$-perturbed
 weights, rather than one single weight. This sequence
 leads to moments, polynomials and a (fuller) matrix evolving
 according to the discrete KP-hierarchy. The associated $\tau$-functions
 have integral, as well as vertex operator representations.
 Among the examples considered, we mention: nested Calogero-Moser
 systems, concatenated solitons and $m$-{\em periodic}
 sequences of weights. The latter lead to $2m+1$-band
 matrices and generalized orthogonal polynomials,
 also arising in the context of a Riemann-Hilbert problem.
 We show the Riemann-Hilbert factorization is tantamount
 to the factorization of the moment matrix into the product
of a lower- times upper-triangular matrix.
\end{abstract}
\tableofcontents{}

\bigbreak

\section{Introduction}

The starting point in the standard theory of orthogonal polynomials
is a single weight $\rho(z)dz$ on an interval of the real line. The
latter leads to moments $\mu_{ij}=\la z^i,z^j \rho\ra $, depending
on $i+j$ only; in turn, moments lead to polynomials $p_n(z)$,
defined by the determinant (0.2) below and the spectral relation
$zp_n=(Lp)_n$ defines tridiagonal semi-infinite matrices $L$. An
important recent development in this ancient theory is that the
perturbed weight $e^{\sum_1^{\iy} t_iz^i}\rho(z)dz$ leads to
$t$-dependent tridiagonal matrices $L(t)$ satisfying the standard
Toda lattice equations; the determinants of the principal minors of
the moment matrix are $\tau$-functions for the Toda lattice and are
representable as integrals over Hermitean matrices, as developped
extensively in \cite{AvM1}.

This paper is designed to show the reader how the introduction of
an infinite family of weights $\rho_j(z)dz$, rather than a family
$z^j \rho(z)dz$ generated by one weight $\rho(z)dz$, leads to a
theory having many features in common with the classic situation
above. The weights lead to ``moments" $\mu_{ij}$, to a
semi-infinite moment matrix $m_{\iy}$, to polynomials $p_n(z)$, as
in (0.2), and to semi-infinite matrices
 $L$ of type (0.4) below, defined by $zp_n(z)=(Lp(z))_n$.
 We mainly deal with:

\noindent{\sl (i)} $t$-deformations ($t=(t_1,t_2,...)$)
 $$ \rho_j(t;z)=
e^{\sum_1^{\iy} t_iz^i}\rho_j(z)dz,~~~~t=(t_1,t_2,...)\in
\BC^{\iy},
 ~z \in\BR,~j=0,1,2,...
$$
of the weights; they imply for the matrix $L$ the so-called
``discrete KP-hierarchy" in $t$; this hierarchy is fully described
in \cite{AvM2}, and a large class of solutions is explained
 in section 1.

Occasionally, shall we deal with

\noindent{\sl (ii)} $(t,s)$-deformations
 \footnote{where the $F_i(t)$ are the elementary Schur polynomials
 $e^{\sum_1^{\iy} t_i z^i}=\sum_{i=0}^{\iy}F_i(t)z^i$}
 $$
\rho_j(t,s;z)=e^{\sum_1^{\iy}t_i z^i}
 \sum_{\ell=0}^{\iy} F_{\ell}(-s)\rho_{j+\ell}(z)
,~~t,s \in \BC^{\iy}, ~z \in \BR,~j=0,1,2,... .
$$
of the weights $\rho_j$; they imply for $L$ the 2d-Toda hierarchy,
as described in \cite{UT,AvM3} and summarized in section 2.

 To be specific, given a family of weights $\rho_0(z) dz,\rho_1(z) dz,...$
on $\BR$, and their $t$-deformations
$$
\rho_j^t(z)dz:=\rho_j (t;z)dz=e^{\sum_1^{\iy} t_k z^k}\rho_j(z)dz,
$$
define the ``moments", with regard to the usual integration in
$\BR$:
\be
\mu_{ij}:=\la z^i,\rho_j(z)\ra\quad\mbox{and}\quad\mu_{ij}(t):=\la
z^i,\rho^t_j(z)\ra,
\ee
and the moment matrix
$$
m_n(t):=\left(\mu_{ij} (t)   \right)_{0\leq i,j \leq n-1}
.
$$
Then the semi-infinite moment matrix $m_{\iy}$
 satisfies the linear differential equations
 $$
 \frac{\pl m_{\iy}}{\pl t_k}=\Lb^k m_{\iy},
 $$
 where $\Lb$ denotes the standard shift matrix.
 They form a infinite set of commuting vector fields.
 Generically the semi-infinite moment matrix $m_{\iy}$
  admits a (unique) factorization into upper- and
  triangular matrices $S_1$ and $S_2$ respectively,
with $S_1$ having $1$'s on the diagonal:
  $$
  m_{\iy}=S_1^{-1} S_2.
  $$

  Consider the vector $p(t,z):=(p_n(t,z))_{n\geq 0}$ of
monic polynomials in $z$, depending\footnote{$\chi(z):=(1,z,z^2,...)$ and
 $\chi^*(z):=\chi(z^{-1})$.} on
$t=(t_1,t_2,...)\in
\BC^{\iy}$,
\be
p_n(t,z):=S_1 \chi(z):=\frac{1}{\det
m_n(t)}\det\left(\begin{array}{ccc|c}
\mu_{00}(t)&...&\mu_{0,n-1}(t)&1\\
\vdots& &\vdots&\vdots\\
\mu_{n-1,0}(t)&...&\mu_{n-1,n-1}(t)&z^{n-1}\\
\hline
\mu_{n0}(t)&...&\mu_{n,n-1}(t)&z^n
\end{array}\right).
\ee
The eigenvalue problem
\be
zp(t,z)=L(t)p(t,z)
\ee
 or, alternatively, the $S_1$-matrix in the
 factorization above, gives rise to the semi-infinite
 matrix
\bea
L&=&S_1 \Lb S_1^{-1}\nonumber\\ &=&\Lb
+a\Lb^0+\Lb^{\top}b+\Lb^{\top 2}c+...=
\left(\begin{array}{cccc}
a_0&1&0&0\\
b_0&a_1 &1 &0 \\
c_0&b_1&a_2&1 \\
\ddots&\ddots&\ddots&\ddots
\end{array}\right).
\eea
The polynomials $p_n(t,z)$ also give rise to a {\em Grassmannian
flag} of nested infinite-dimensional planes $...\supset
 \WR^t_n \supset \WR^t_{n+1} \supset ...$ ,
given by\footnote{$\HR_+:=~\mbox{span}~\{1,z,z^2,...\}$.}
\be
\HR_+ \supset \WR^t_n:=\WR_n e^{-\sum_1^{\iy} t_i z^i}:=\Span
\left\{  p_n(t,z),p_{n+1}(t,z)... \right\}.
\ee
We shall also need the associated ``Vandermonde"
determinants\footnote{$\Dt_n(z)=\prod_{1\leq j <i
\leq n}(z_i-z_j)$},
\be
\Delta_n^{(\rho)}(z)=\det\left(\rho_{\ell-1}(z_k)\right)_{1\leq\ell,k\leq
n}, ~~\Delta_n(z)=\det(z_k^{\ell-1})_{1\leq\ell,k\leq n},
\ee
and the simple vertex operator,
\be
X(t,z):=e^{\sum_1^{\iy}t_iz^i}e^{-\sum_1^{\iy}
\frac{z^{-i}}{i}\frac{\pl}{\pl t_i}};
\ee
this is, in disguise, a Darboux transform acting on KP
$\tau$-functions. We now state:

\begin{theorem} Given the moments (0.1) and the construction above, the semi-infinite
matrix $L$ in (0.4) satisfies the \underline{discrete KP hierarchy}
\be
\frac{\pl L}{\pl t_n}=[(L^n)_+,L],\quad n=1,2,...,
\ee
and has the following $\tau$-function representation\footnote
 {where $F_{\ell}(\tilde \pl)=F_{\ell}(\frac{\pl}{\pl t_1},
 \frac{1}{2}\frac{\pl}{\pl t_2},
 \frac{1}{3}\frac{\pl}{\pl t_3},...)$, for the elementary
 Schur polynomials $F_{\ell}$. The symbol
 $F_{\ell}(\tilde \pl)f\circ g$ is the customary
 Hirota symbol.}
\be
L=\sum_{\ell=0}^{\iy}\mbox{diag}\left(\frac{F_{\ell}(\tilde\pl)
\tau_{n+2-\ell}\circ\tau_n}
{\tau_{n+2-\ell} \tau_n}\right)_{n \geq 0 }\Lb^{1-\ell}
\ee
in terms of a sequence of $\tau$-functions $(\tau_0=1,\tau_1,
\tau_2,...)$, which enjoys many different representations:
\bea
\tau_n(t)&=&\det\left(\mu_{\ell,k}(t)\right)_{0\leq\ell,k\leq
n-1}\mbox{ \hspace{3.5cm} (moment representation)}\nonumber\\
 & & \\
&=&\frac{1}{n!}\int
...\int_{\BR^n}\Delta_n(z)\Delta_n^{(\rho)}(z)\prod_{k=1}^n
\left(e^{\sum t_iz^i_k}dz_k\right)\mbox{\hspace{0cm} (integral
representation)}\nonumber   \\
 & & \\
 &=&\det \left(
\mbox{Proj}:~ e^{-\sum_1^{\iy}t_iz^i}z^{-n}\WR_n
\rightarrow
\HR_+\right) \mbox{\hspace{1.5cm} (flag representation)}\nonumber
\\
& & \\
 &=&\det \left(
\mbox{Proj}:~ e^{\sum_1^{\iy}t_iz^i}z^{n}\WR^*_n
\rightarrow
\HR_+\right)
\mbox{\hspace{1.1cm} (dual flag representation)}
 \nonumber\\
& & \\
 &=& \left(\int_{\BR}~dz
~z^{n-1}   \rho_{n-1}(z) X(t,z)
\right)
\tau_{n-1}(t)\mbox{\hspace{1cm} (vertex representation)}
 \nonumber \\
& &
\eea
where
\bea
\WR_n e^{-\sum_1^{\iy} t_i z^i}&=&\Span
\left\{  p_n(t,z),p_{n+1}(t,z)... \right\}\subset \HR_+
\nonumber\\
\WR_n^{\ast }~e^{\sum_1^{\iy} t_i z^i}&=&\Span
\left(\left\{\int_{\BR}\frac{\rho^t_j(u)du}{z-u},j=0,...,
n-1\right\}\oplus \HR_+\right)\supset \HR_+.\nonumber\\
\eea
 The polynomials (0.2) have the following
representations
\bea
p_n(t,z)& =&    \frac{\det \left(z\mu_{ij}(t)-\mu_{i+1,j}(t)
\right)_{0\leq i,j\leq n-1}}
{\det (\mu_{ij}(t))_{0\leq i,j\leq n-1}}\\ &=&
\frac{1}{n!\tau_n(t)} \int ...\int_{\BR^n}\Delta_n(z)
\Delta_n^{(\rho)}(z)\prod^n_{k=1}
\left(e^{\sum
t_iz^i_k}\left(z-z_k \right)dz_k\right),\nonumber\\
\eea
and satisfy the eigenvalue problem $Lp=zp$.
\end{theorem}

Notice that formulae (0.16) and (0.17) go in parallel with (0.10)
and (0.11). Formula (0.17) is a generalization of a formula for
classical orthogonal polynomials already appearing last century in
the work of Heine \cite{H}.

\bigbreak

We shall apply this theorem to a variety of examples, corresponding
to sections 4 to 8 ($\delta (x)$ is the customary delta-function, i.e.,
$\int_{\BR}\delta(x)f(x)dx=f(0)$) :

\begin{eqnarray*}
\rho_j(z):=z^j\rho(z)& &
~~~\mbox{{\em tridiagonal matrix $L$}}\\
\rho_{j+km}(z):=z^{km}\rho_j(z) & &
~~~\mbox{{\em $2m+1$-band matrix $L^m$ }}\\
\rho_k(z)=\delta(z-p_{k+1})-\lb^2_{k+1}\delta(z-q_{k+1})
& &~~~\mbox{{\em concatenated solitons}}\\
\rho_k(z)=\dt'(z-p_{k+1})+\lb_{k+1}\dt(z-p_{k+1})&&~~~
\mbox{{\em nested Calogero-Moser systems}}
 \\
\rho_k(z)=(-1)^k\dt^{(k)}(z-p)-\dt^{(k)}(z+p)
& &~~~\mbox{{\em upper-triangular $L^2$}}.
\end{eqnarray*}

\vspace{1cm}

The \underline{first example} leads to the standard Toda lattice
and the the classic theory of {\em orthogonal polynomials}. Since
the work of Fokas, Its and Kitaev \cite{FIK}, the {\em
Riemann-Hilbert
 method} is a
device to obtain asymptotics for orthogonal polynomials; for
semi-classical asymptotics, see Bleher and Its \cite{BI}. We show
 \be
\mbox{Riemann-Hilbert factorization}~~~\Longleftrightarrow
~~~ \mbox{factorization}~~m_{\iy}=S_1^{-1} S_2.
\ee
To be precise, we show the Riemann-Hilbert matrices $Y_n$ take on
the following form ($\chi(z)$ and $\chi^*(z)$ are as in footnote 2
and $h_{n-1}:=\tau_{n}/\tau_{n-1}$):

\bea
Y_n(z)&=&\pmatrix{\left(S_1 \chi(z) \right)_{n} &
\frac{1}{z}\left(S_2 \chi^*(z) \right)_{n} \cr \cr
h_{n-1}^{-1}\left(S_1 \chi(z) \right)_{n-1} &
 h_{n-1}^{-1}\frac{1}{z}\left(S_2 \chi^*(z) \right)_{n-1}
 \cr
 }\nonumber\\&&\nonumber\\&=&
\pmatrix{\frac{\tau_{n}(t-[z^{-1}])} {\tau_n(t) } z^{n}&
\frac{\tau_{n+1}(t+[z^{-1}])}{\tau_n(t)}
z^{-n-1}  \cr
\frac{\tau_{n-1}(t-[z^{-1}])}
{\tau_{n}(t)} z^{n-1}&
\frac{\tau_{n}(t+[z^{-1}])}{\tau_{n}(t) }
z^{-n} \cr }.
\eea

The \underline{second example}, which is novel and which is
developped in section 5, involves a finite set of weights
$$\rho_0(z)dz,...,\rho_{m-1}(z)dz$$ for $m\geq 2$, which we extend
into an infinite ``{\em $m$-periodic}" sequence
$$\rho_0(z)dz,...,\rho_{m-1}(z)dz,z^m\rho_0(z)dz,...,z^m\rho_{m-1}(z)dz
,\hspace{3cm}$$
$$\hspace{5cm}z^{2m}\rho_0(z)dz,...,z^{2m}\rho_{m-1}(z)dz,...~.$$
This sequence leads naturally to {\sl generalized orthogonal
polynomials} $p_n(z)$ by the recipe (0.2), which enjoys the
following properties:

 \medbreak

\noindent
$\left\{
\begin{tabular}{l}
(i) the polynomials $p_n(z)$ satisfy the {\em orthogonality
relations}
 \\ ~~~~~~ $\la p_i(z),
\rho_j(z) \ra=0$ for $i \geq j+1$;      \\
 (ii) Applying $z^m$ to the vector
 $p(z):=(p_0(z),p_1(z),...)$
 leads to a \\ ~~~~~$2m+1$-band matrix $L^m$. \\
 (iii) The $t$-evolution
 $e^{\sum_1^{\iy}t_iz^i}\rho_k(z)$ implies $L$ evolves
 according to the \\ ~~~~~~discrete KP hierarchy.   \\
\end{tabular}
\right.
$

The discrete KP-hierarchy on $2m+1$-band matrices has been
 studied in \cite{vMM}; see also \cite{G}. We also
 formulate here a Riemann-Hilbert problem, which
 should characterize the generalized orthogonal
 polynomials.

Another interesting set of examples is provided by picking as
weights various combinations of (standard) $\dt$-functions, which
lead to concatenated soliton solutions, Calogero-Moser systems,
etc...

We wish to thank Leonid Dickey for insightful comments and
 criticism, which lead to recognize the importance of
 the integral in
(0.15) beyond its formal aspects. L. Dickey has shown in a very
 interesting recent paper \cite{Dickey} that the discrete KP
 hierarchy is the most natural generalization of the modified KP. We also
 thank Alexander Its and Pavel Bleher for having
 explained to us the Riemann-Hilbert
 problem, and for having posed the problem of
 finding the connection with the matrix factorization
 of the moment matrix. We also thank Taka Shiota for a number of
interesting conversations.

\section{Vertex operator solutions to the discrete KP hierarchy}

In \cite{AvM2}, we discussed the discrete KP hierarchy and found a
general method for generating its solutions, in both, the bi- and
semi-infinite situations; this paper mainly deals with the
semi-infinite case. In \cite{AvM2} and \cite{AHV}, we gave an
application of the bi-infinite discrete KP to the $q$-KP equation.
In general, the main features are summarized in the following
statement, whose proof can be found in \cite{AvM2}:

\begin{theorem} From an arbitrary KP $\tau$-function and a
sequence of real functions
$(...,\nu_{-1}(\lb),\nu_{0}(\lb),\nu_{1}(\lb),...)$, defined on
$\BR$, one constructs the infinite sequence of $\tau$-functions:
$\tau_0=\tau$ and, for $ n> 0$,
\bea
\tau_0(t)&=&\tau(t)\nonumber\\
\tau_{ n}(t)& = &\left(\int X (t,\lambda) \nu_{n-1}(\lb)d\lb~...\int
X (t,\lambda)\nu_0(\lambda)d\lb\right)
\tau (t),~~n> 0\nonumber\\
\tau_{-n}(t)&=&\left(\int X (-t,\lambda) \nu_{-n}(\lb)d\lb~...\int X
(-t,\lambda)\nu_{-1}(\lambda)d\lb\right) \tau (t),~~~~n>
0.\nonumber\\
\eea
Then the bi-infinite vector
\be
\Psi(t,z)=\left( \frac{\tau_n(t-[z^{-1}])}{\tau_n(t)}
e^{\sum^{\iy}_1 t_iz^i} z^n   \right)_{n\in\BZ}
\ee
and bi-infinite matrix
\be
L=\sum_{\ell=0}^{\iy}\mbox{diag}~\left(\frac{F _{\ell}(\tilde\pl)
\tau_{n+2-\ell}\circ\tau_n}
{\tau_{n+2-\ell} \tau_n} \right)_{n \in \BZ}\Lb^{1-\ell}
\ee
satisfy the discrete KP-hierarchy equations for $n=1,2,...$:
\be
 \frac{\pl\Psi}{\pl
t_k}=(L^k)_+\Psi~~~\mbox{and}~~~ \frac{\pl L}{\pl t_k}=[(L^k)_+,L],
~~\mbox{ with}~~ L\Psi(t,z)=z\Psi(t,z).
\ee
Then $\tau_n(t)$ is given by the following projection
\be
\tau_n(t)=\det\left(\mbox{Proj\,} : e^{-\sum^{\iy}_1
t_iz^i}z^{-n}\WR_n\lrg\HR_+\right),
\ee
where the Grassmannian flag $...\supset \WR_n \supset
\WR_{n+1}\supset...$ is given by
\be
\WR_n:=\mbox{span}_{\BC}\{\Psi_n(t,z),\Psi_{n+1}(t,z),...\}.
\ee
Conversely, a Grassmannian flag $...\supset \WR_n \supset
\WR_{n+1}\supset...$, given by (1.6), with functions $\Psi_n(t,z)$
satisfying the asymptotics $\Psi_n(t,z)=e^{\sum
t_iz^i} z^n(1+O(1/z))$ leads to the discrete KP-hierarchy.
\end{theorem}

\vspace{0.5cm}

\noindent \underline{\em Remark}~: A semi-infinite discrete
KP-hierarchy with
$\tau_0(t)=1$ is equivalent to a bi-infinite discrete
KP-hierarchy with
$\tau_{-n}(t)=\tau_n(-t)$ and $\tau_0(t)=1$; in the above
theorem, this amounts to setting $\tau_0(t)=1$ and
$\nu_{-n}(\lb):=\nu_{n-1}(\lb),~n=1,2,...$. We extend the
semi-infinite flag $\WR_{0}=\HR_+ \supset...\supset \WR_n \supset
\WR_{n+1}\supset...$, by setting
$\WR_{-n}=\WR^{\ast}_n$, for $n \geq 0$.

\bigbreak

\section{Moment matrix factorization and solutions to
discrete KP and 2d-Toda }

In (0.1), we considered $t$-deformations of the sequence of
 weights, with $t\in \BC^{\iy}$,
$$
(\rho^t_0(z),\rho^t_1(z),...)~,~~t\in
\BC^{\iy}~\mbox{with}~~\rho^t_j(z)=e^{\sum_0^{\iy}t_k z^k}\rho_j(z).
$$
 As announced in the introduction, we consider further
 deformations of the sequence of weights, in the $(t,s)$-direction,
 \be
\rho_j(t,s;z)=e^{\sum_1^{\iy}t_i z^i}
 \sum_{\ell=0}^{\iy} F_{\ell}(-s)\rho_{j+\ell}(z)
,~~t,s \in \BC^{\iy}, ~z \in \BR,~j=0,1,2,... .
\ee
and the corresponding moment matrix
\be
m_{n}(t,s)=(\mu_{ij}(t,s))_{0\leq i,j < n-1},~\mbox{with}
~~\mu_{ij}(t,s)=\la z^i,\rho_j(t,s;z)\ra.
\ee
We now state the following proposition (e.g., see
\cite{AvM3,AvM4}):

\begin{proposition}
The matrix $m_{\iy}(t,s)$ satisfies the differential equations
\be
\frac{\pl m_{\iy}}{\pl t_k}=\Lb^k m_{\iy}~,~~~
\frac{\pl m_{\iy}}{\pl s_k}=-m_{\iy} \Lb^{k \top}
.\ee
Factorizing the matrix $m_{\iy}(t,s)$ into the product of lower-
and upper-triangular matrices $S_1$ and $S_2$, with $S_1$ having
$1$'s along the diagonal:
\be
m_{\iy}(t,s)=  S_1^{-1}(t,s) S_2(t,s),
\ee
the sequence of wave functions\footnote{
 $\xi_1(z):=\sum t_i z^i$ and
 $\xi_2(z):=\sum s_i z^{-i}$; also $\chi(z):=(1,z,z^2,...)$ and
 $\chi^*(z):=\chi(z^{-1})$.}, derived from
 $S_1$ and $S_2$,
\be
\Psi_i(t,s;z)=e^{\xi_i(z)}S_i\chi(z)~~~
\Psi_i^*(t,s;z)=
e^{-\xi_i(z)}(S_i^{\top})^{-1}\chi^{\ast}(z),
\ee
can be expressed in terms of $\tau$-functions
 $\tau_n(t,s)=\det m_n$, as follows:
\begin{eqnarray}
\Psi_1(t,s;z)&=&\biggl(
        {\tau_n(t-[z^{-1}],s)\over\tau_n(t,s)}
e^{\sum^{\iy}_1 t_iz^i}z^n
\biggr)_{n\in\BZ}\nonumber\\
\Psi_2(t,s;z)&=&\biggl(
        {\tau_{n+1}(t,s-[z])\over\tau_n(t,s)}
e^{\sum^{\iy}_1 s_iz^{-i}}z^n
\biggr)_{n\in\BZ}\nonumber\\
\Psi^*_1(t,s;z)&=&\Biggl(\frac{\tau_{n+1}(t+[z^{-1}],s)}
{\tau_{n+1}(t,s)}e^{-\sum^{\iy}_1
t_iz^i}z^{-n}\Biggr)_{n\in\BZ}\nonumber\\
\Psi^*_2(t,s;z)&=&\Biggl(\frac{\tau_n(t,s+[z])}
{\tau_{n+1}(t,s)}e^{-\sum^{\iy}_1
s_iz^{-i}}z^{-n}\Biggr)_{n\in\BZ},\nonumber\\
\end{eqnarray}
with $\Psi_i(t,s)$ satisfying the following differential
equations\footnote{$A_+$ and $A_-$ denote the upper-triangular
 and strictly lower-triangular part of the matrix $A$,
 respectively.}
$$
\frac{\pl \Psi_i}{\pl t_n}=(L_1^n)_+\Psi_i~,~~
 \frac{\pl \Psi_i}{\pl s_n}=(L_2^n)_-\Psi_i
 ~\mbox{with}~L_1=S_1\Lb S_1^{-1},~
 L_2=S_2\Lb^{-1} S_2^{-1}.
$$
The $\tau$-functions satisfy bilinear identities, for all $n,m \geq
0$,
$$
  \oint_{z=\iy}\tau_n(t-[z^{-1}],s)\tau_{m+1}(t'+[z^{-1}],s')
  e^{\sum_1^{\iy}(t_i-t'_i)z^i}
  z^{n-m-1}dz \hspace{2cm}
$$
  \be
  =\oint_{z=0}\tau_{n+1}(t,s-[z])\tau_m(t',s'+[z])
  e^{\sum_1^{\iy}(s_i-s'_i)z^{-i}}z^{n-m-1}dz,
  \ee
  and therefore the KP hierarchy in each of the
  variables $t$ and $s$.
\end{proposition}

\bigbreak

The following corollary can be found in \cite{AvM4}:

 \begin{corollary} 2d-Toda $\tau$-functions satisfy the
 following (Fay-like) identities for arbitrary $z,u,v\in \BC$.
  $$
  \tau_n(t-[z^{-1}],s+[v]-[u])\tau_n(t,s)-\tau_n(t,s+[v]-[u])\tau_n(t-
  [z^{-1}],s)$$
  \be=\frac{v-u}{z}\tau_{n+1}(t,s-[u])\tau_{n-1}(t-[z^{-1}],s+[v]).
  \ee
  \end{corollary}

\bigbreak

Introduce now the residue pairing about $z=\iy$,
 between $f=\sum_{i\geq 0}a_i z^i \in
\HR^+$ and $g =\sum_{j \in \BZ}b_j z^{-j-1}\in \HR$:
\be
\la f,g\ra_{\iy}=\oint_{z=\iy}f(z)g(z)\frac{dz}{2 \pi i}=
\sum_{i \geq 0}a_i b_i
 ,
\ee
where the integral is taken over a small circle
 about $z=\iy$.

 But setting $s=s'$ and $m\leq n-1$, the right hand integrand
 of (2.7) is holomorphic and so the right hand side of (2.7)
 vanishes. Of course, freezing $s=s'$ yields the discrete KP-hierarchy;
 see \cite{AvM2}.
 Therefore
\be
\oint_{z=\iy}\tau_n(t-[z^{-1}],s)
 \tau_{m+1}(t'+[z^{-1}],s)
  e^{\sum_1^{\iy}(t_i-t'_i)z^i}
  z^{n-m-1}dz=0~\mbox{for}~~n \geq m+1,
  \ee
and so for $n \geq m+1$,
\be
\oint_{z=\iy}e^{\sum_1^{\iy}t_i z^i}
z^n\frac{\tau_n(t-[z^{-1}],s)}{\tau_n(t,s)}
 e^{-\sum_1^{\iy} t'_i z^i}z^{-m-1}\frac{\tau_{m+1}(t'+[z^{-1}],s)}
 {\tau_m(t,s)}
dz=0.
  \ee
Defining the linear space $\WR_n^*$ as the space of functions
perpendicular to $\WR_n$ for the residue pairing (2.9), we thus
have for fixed $s$, by virtue of (1.6), (2.6) and then (2.11),
\bea
\WR^t_n&=&\Span \{ z^j \frac{\tau_j(t-[z^{-1}],s)}
{\tau_j(t,s)}, ~~j\geq n  \}=e^{-\sum t_i z^i}\WR_n\nonumber\\
\WR^{* t}_n&=&\Span \{ z^{-j} \frac{\tau_j(t+[z^{-1}],s)}
{\tau_{j-1}(t,s)}, ~~j\leq n  \}=e^{\sum t_i z^i}\WR^*_n.
\eea
It also shows that $\tau_n(t,s)$ can be obtained from those spaces
in two different ways (for fixed $s$):
\bea
\tau_n(t,s)&=&\det\left(\mbox{Proj\,} : e^{-\sum
t_iz^i}z^{-n}\WR_n\lrg\HR_+\right)\nonumber\\
&=&\det\left(\mbox{Proj\,} : e^{\sum
t_iz^i}z^n\WR_n^*\lrg\HR_+\right),
\eea
where the multiplication by $z^{-n}$ and $z^n$ makes the
corresponding linear spaces have ``genus zero", in accordance with
the terminology of Segal-Wilson \cite{SW}.

\bigbreak

As a {\em special case (H\"ankel matrices)} , consider the sequence
of weights
\be
\rho_j(z)dz=z^j \rho(z)dz.
\ee
Then the $(t,s)$-deformations take on the following form:
$$
\rho_j(t,s;z)=e^{\sum t_iz^i}\sum_{\ell \geq 0}
F_{\ell }(-s)z^{\ell+j}\rho(z)
 =e^{\sum (t_i-s_i) z^i}z^j \rho(z),
$$
thus depending on $t-s$ only. Therefore $\mu_{ij}(t,s)$ depends
only on $t-s$ and $i-j$ ($m_{\iy}$ is a H\"ankel matrix) and so
$\tau_n(t,s)$ depends only on $t-s$. Therefore, in this case we may
replace $t-s$ by $t$.

In this case, the matrix $m_{\iy}$ is symmetric, which simplifies
the factorization (2.4) above. Indeed:
\be
m_{\iy}(t)=S_1^{-1}S_2=S_1^{-1}hS_1^{-1 \top}=S^{-1}(t) S^{\top
-1}(t),
\ee
 upon setting
 \be
 S=h^{-1/2}S_1=h^{1/2}S_2^{-1 \top}.
 \ee

\section{Weights, flags and dual flags}

\setcounter{equation}{0}

The purpose of this section is to prove Theorem 0.1. The point is
to derive the $\tau$-functions from the Grassmannian flag (1.6).
Unfortunately, the matrix associated with the projection (1.5) is
{\em infinite}; therefore taking its determinant would be
non-trivial, although possible. However, it turns out to be
infinitely easier to consider the {\em dual flag}, which leads to a
{\em finite projection matrix}, whose determinant is the same
$\tau$-function.

To carry out this program, we equip the space $\HR:=\Span\{ z^i,~i
\in
\BZ\}$ with two inner products: the usual
one
\be
\la f,g\ra =\int_{\BR}f(z)g(z)\,dz,
\ee
and remember the residue pairing about $z=\iy$, between
$f=\sum_{i\geq 0} a_i z^i
\in
\HR^+$ and $g =\sum_{j \in \BZ}b_j z^{-j-1}\in \HR$:
\be
\la f,g\ra_{\iy}=\oint_{z=\iy}f(z)g(z)\frac{dz}{2 \pi i}=
\sum_{i \geq 0}a_i b_i
 ,
\ee
where the integral is taken over a small circle about $z=\iy$. The
two pairings, which will be instrumental in linking the flag to the
dual flag, are related as follows:

\begin{lemma}
\be
\la f,g\ra =\left\la f,\int_{\BR}\frac{g(u)}{z-u}du\right\ra_{\iy}.
\ee
\end{lemma}

\proof
Expanding the integral above into an asymptotic series, which we
take as its definition,
 \bea
\int_{\BR}\frac{g(u)}{z-u}du
&=&\frac{1}{z}\int_{\BR} g(u)\sum_{j \geq 0} \left(\frac{u}{z}
\right)^j du
\nonumber\\
&=&\frac{1}{z}\sum_{j \geq 0} z^{-j}\int_{\BR} g(u) u^j du,
\eea
we check that for holomorphic functions $f$ in $\BC$,
\bea
\left\la f,\int_{\BR}\frac{g(u)}{z-u}du\right\ra_{\iy}.
&=&\left\la \sum_{i\geq 0} a_i z^i,
 \frac{1}{z}\sum_{j \geq 0} z^{-j}\int_{\BR} g(u) u^j du,
\right\ra_{\iy}.
\nonumber\\
&=& \sum_{i\geq 0} a_i\int_{\BR} g(u)u^i du
 \nonumber\\
 &=& \int_{\BR} g(u)\sum_{i\geq 0} a_i u^i du
 \nonumber\\
 &=& \la f,g\ra.
 \eea
\qed

\bigbreak

\remark The series (3.4) only converges outside the support of
 $g(u)$. So, in general, the series (3.4) diverges, even for large
 $z$. In specific examples, this integral will have a precise
 meaning; see sections 4 and 5.

\bigbreak

To the family of functions $\rho_0(z),\rho_1(z),...$ on $\BR$, and
$\rho_j^t(z):=e^{\sum t_kz^k}\rho_j(z)$, we associate the flag of
spaces $\WR_{0}=\HR_+ \supset...\supset \WR_n \supset
\WR_{n+1}\supset...$, defined by
\begin{eqnarray}
\WR_n&:=&\Bigl(\Span\{\rho_1,\rho_1,...,\rho_{n-1}\}\Bigr)^{\bot}\nonumber\\
 &=& \{f\in\HR_+\mbox{\,\,such that
$\la f,\rho_i\ra =0,~0 \leq i \leq n-1\}$}
\end{eqnarray}
with respect to the inner product (3.1). So, throughout we shall be
playing with the following two representations of the moments:
\be
\mu_{ij}=\la z^i~,~ \rho_j^t(z) \ra=\left\la
z^i~,~\int_{\BR}\frac{\rho^t_j(u) du}{z-u}
\right\ra_{\iy}
\ee
With the moments
$
\mu_{ij}(t):=\la z^i,\rho^t_j\ra,
$
we associate the monic polynomials $p_k(t,z)$ in $z$ of degree $k$,
introduced in (0.2). As usual, set $$\WR_n^t=e^{-\sum t_i z^i}\WR_n
~~\mbox{and its dual}~~ \WR_n^{\ast t}=e^{\sum t_i
z^i}\WR^{\ast}_n.$$
 As we showed in (2.12), for the residue pairing we have:
$$
\left\la
\WR_n^t, \WR^{\ast t}_n
\right\ra_{\iy}=
\left\la
\WR_n, \WR^{\ast}_n
\right\ra_{\iy}=0.
$$
The integral representation (3.9) below of the dual flag already
appears in the work of Mulase \cite{M}, for the case $\rho_j(z)=z^j \rho(z)$.
\bigbreak

\begin{proposition} The flag $\HR_+\supset \WR_1 \supset \WR_2
\supset...$, defined by (3.6) at $t=0$, evolves into
\be
\WR^t_n=(\Span\{\rho^t_0,\rho^t_1,...,\rho^t_{n-1}\})^{\bot}=
\Span\{p_n(t,z),p_{n+1}(t,z),...\} \subset \HR_+,
\ee
and the dual flag $\HR_+\subset \WR^{\ast}_1 \subset \WR^{\ast}_2
\subset...$, evolves into
\be
\WR_n^{\ast t}=\Span\left(\left\{\int\frac{\rho^t_j(u)du}{z-u},j=0,...,
n-1\right\}\oplus \HR_+\right).
\ee
\end {proposition}

\proof Indeed to show (3.8), it suffices to check the following, for $k\geq j+1$
and the polynomials
$p_k(t,z)=\frac{1}{a_{kk}(t)}\displaystyle{\sum^k_{i=0}}a_{ki}(t)z^i$,
defined in (0.2):
\bea
\la
p_k(t,z),\rho^t_j\ra&=&\frac{1}{a_{kk}(t)}\sum_{i=0}^ka_{ki}(t)\la
z^i,\rho^t_j\ra\nonumber\\
&=&\frac{1}{a_{kk}(t)}\sum_{i=0}^ka_{ki}\mu_{ij}(t)\nonumber\\
&=&\frac{1}{a_{kk}(t)}\det\left(\begin{array}{ccc|c}
\mu_{00}(t)&...&\mu_{0,k-1}(t)&\mu_{0j}(t)\nonumber\\
\vdots& &\vdots&\vdots\\
\hline
\mu_{k0}(t)&...&\mu_{k,k-1}(t)&\mu_{kj}(t)
\end{array}\right)=0.\\
\eea
To prove the dual statement (3.9), one checks for $k\geq j+1$
$$
\left\la p_k(t,z),\int_{\BR}\frac{\rho^t_j(u)du}{z-u}\right\ra_{\iy}=\la
p_k(t,z),\rho^t_j(z)\ra=0,
$$
using Lemma 3.1, and, of course,
$$
\left\la
p_k(t,z),z^{\ell}\right\ra_{\iy}
=0,~~\mbox{for all}~k,\ell
\geq 0.
$$
\qed

\bigbreak
Remember from (2.13), the $\tau$-functions $\tau_n(t)$ can be
computed in two different ways:
\bea \tau_n(t)&=&\det\left(\mbox{Proj\,} : e^{-\sum
t_iz^i}z^{-n}\WR_n\lrg\HR_+\right)\nonumber\\
&=&\det\left(\mbox{Proj\,} : e^{\sum
t_iz^i}z^n\WR_n^*\lrg\HR_+\right). \eea
  We shall need
the following lemma concerning Vandermonde-like
determinants, extending a lemma mentioned in \cite{M}:

\begin{lemma}
\be
\sum_{\sg\in\prod}\det\left(u_{\ell,\sg(k)}v_{k,\sg(k)}
\right)_{1\leq\ell,k\leq
n}=\det\left(u_{\ell,k}\right)_{1\leq\ell,k\leq n}
\det\left(v_{\ell,k}\right)_{1\leq\ell,k\leq n}.
\ee
\end{lemma}

\vspace{0.8cm}

\noindent{\underline{\em Proof of theorem 0.1}}: Since $z^n
\WR_n^{\ast} \supset z^n\HR_+$, the matrix of the projection (3.9)
onto $\HR_+$, involving $\WR_n^{\ast}$, reduces to a {\em finite}
matrix, whereas the projection involving $\WR_n$ would involve an
{\em infinite} matrix! This is the point of using $\WR^{\ast}_n$
rather then $\WR_n$. Therefore the matrix of the projection
$$\mbox{Proj}:~e^{\sum t_k z^k}  z^n
\WR_n^{\ast t}\lrg \HR_+$$
is obtained by putting all coefficients of
$$
e^{\sum t_k z^k}  z^n \int\frac{\rho_j(u) du}{z-u} ~~\mbox{for}~~
(0\leq j \leq n-1)  ~~\mbox{and}~~      e^{\sum t_k z^k}  z^{n+j} ~~\mbox{for}~~
(0\leq j < \iy)
$$
in the $j$th and $n+j$th columns respectively, starting on top with
$z^0,z^1,...$. Since for any power series
$$
z^j\mbox{-coef of}~f=\oint_{z=\iy} z^{-j-1} f(z) \frac{dz}{2 \pi
i}=\la z^{-j-1}, f(z)\ra_{\iy},
$$
we have
\be
\tau_n(t)=\det\left(\begin{array}{c|c}
A&0\\
\hline
B&C
\end{array}\right)\,=\det A \det C\,=\det A,
\ee
 where
\bea C&=&\left(  \mbox{coef}_{z^{n+i}} z^{n+j} e^{\sum t_k z^k}
\right)_{0\leq i,j < \iy}\nonumber\\
&=&\MAT{4}1&0&0&...\\
F_1&1&0&...\\
F_2&F_1&1&...\\
\vdots&\vdots&\vdots&\ddots\mat,\nonumber
\eea
and
\bea
A&=&\left(  \mbox{coef}_{z^i}\left( z^n e^{\sum t_k z^k} \int_{\BR}
\frac{\rho_j(u) du}{z-u}
\right)
\right)_{0\leq i,j \leq n-1}\nonumber\\
&=& \left(~ \left\la    z^{n-i-1}e^{\sum t_k z^k}, \int_{\BR}
\frac{\rho_j(u) du}{z-u} \right\ra_{\iy} ~\right)_{0\leq i,j \leq
n-1}\nonumber\\ &=& \left( ~\left\la    u^{n-i-1},e^{\sum t_k
u^k}\rho_j(u)
\right\ra ~\right)_{0\leq i,j \leq n-1}\nonumber\\
&=& \left(  \mu_{n-i-1,j}(t)\right)_{0\leq i,j \leq n-1},
\eea
which provides the $A$-matrix in (3.13), thus establishing (0.10).
Hence,
\begin{eqnarray*}
\tau_n(t)&=&\det\left(\mu_{\ell k}(t)\right)_{0\leq\ell,k\leq n-1}\\
&=&\det\left(\int_{\BR}
z^{\ell}\rho_k(t,z)dz\right)_{0\leq\ell,k\leq n-1}\\
&=&\det\left(\int_{\BR} z^{\ell
-1}_{\sg(k)}\rho_{k-1}(t,z_{\sg(k)})dz_{\sg(k)}\right)_{1\leq\ell,k\leq
n}\mbox{\,\,for a fixed permutation $\sg$}\\
&=&\int ...\int_{\BR^n}\det\left(z^{\ell-1}_{\sg(k)}
\rho_{k-1}(t,z_{\sg(k)})\right)_{1\leq\ell,k\leq n}dz_1...dz_n\\
&=&\frac{1}{n!}\sum_{\sg}\int
...\int_{\BR^n}\det\left(z^{\ell-1}_{\sg(k)}
\rho_{k-1}(t,z_{\sg(k)})\right)_{1\leq\ell,k\leq n}dz_1...dz_n\\
&=&\frac{1}{n!}\int ...\int_{\BR^n}\det(z_k^{\ell-1})_{1\leq
k,\ell \leq n}\det\left(\rho_{\ell-1}(z_k)\right)_{1\leq k,\ell\leq
n}\prod^n_{k=1}
\left(e^{\sum t_iz^i_k}dz_k\right),
\end{eqnarray*}
using Lemma 3.3; this establishes (0.11). Furthermore, we have,
continuing the identities above, that\footnote{using in the fifth
identity $e^{-\sum_1^{\iy} a^i/i} =1-a$.  }
\begin{eqnarray*}
\tau_n(t)&=&\frac{1}{n!}\int
...\int_{\BR^n}
\Delta_n(z)\sum_{\sigma}(-1)^{\sigma}\prod_{\ell=1}^n
\rho_{\ell-1}(z_{\sigma(\ell)})\prod_{k=1}^n\left(e^{\sum
t_iz^i_k}dz_k\right)\\
&=&\frac{1}{n!}\int ...\int_{\BR^n}
\sum_{\sigma}\Delta_n(\sigma^{-1}z)(-1)^{\sigma}\prod_{\ell=1}^n
\rho_{\ell-1}(z_{\ell})\prod_{k=1}^n\left(e^{\sum
t_iz^i_k}dz_k\right)\\
&=&\frac{1}{n!}\int ...\int_{\BR^n}
n!\Delta_n(z) \prod_{\ell=1}^n
\left(\rho_{\ell-1}(z_{\ell})e^{\sum
t_iz^i_{\ell}}dz_{\ell}\right)\\
&=&\int_{\BR}
z^{n-1}\prod_1^{n-1}\left(1-\frac{z_i}{z}\right) e^{\sum
t_iz^i}\rho_{n-1}(z)~ dz \\ & &~~~~~~\int_{\BR^{n-1}}
\Dt_{n-1}(z_1,...,z_{n-1})\prod_{\ell=1}^{n-1}\left(
\rho_{\ell-1}(z_{\ell})e^{\sum t_iz^i_{\ell}}dz_{\ell} \right)\\
&=&\int_{\BR}~dz~ z^{n-1}  \rho_{n-1}(z)e^{\sum t_iz^i}   e^{-\sum
\frac{1}{iz^i}\frac{\pl}{\pl t_i}}\\ &
&~~~~~~\int_{\BR^{n-1}}\Dt_{n-1}(z_1,...,z_{n-1})\prod_{\ell=1}^{n-1}\left(
\rho_{\ell-1}(z_{\ell})e^{\sum t_iz^i_{\ell}}dz_{\ell} \right)\\
&=& \left(\int_{\BR}~dz~ z^{n-1}   \rho_{n-1}(z) X(t,z)  \right)
\tau_{n-1}(t),
\end{eqnarray*}
proving (0.14). Therefore, the sequence $\tau_n(t)$ satisfies (1.1)
in Theorem 1.1 with $\nu_n(z)=z^n \rho_n(z)$ and $\tau_0(t)=1$. The
$\tau$-functions lead to the expression (1.3) for $L$ and to the
expression (1.2) for $\Psi$, which both satisfy the discrete KP
hierarchy, according to Theorem 1.1. Notice that, from (0.15),
(1.5) and (1.2), we have
\be
p_n(t,z)=e^{-\sum_1^{\iy}t_i z^i}\Psi_n(t,z);
\ee
therefore $L$ defined by $\tau$-functions (1.3) agrees with the
semi-infinite $L$, defined by the semi-infinite polynomial
relations $zp(t,z)=Lp(t,z)$, yielding (0.8) and (0.9) for this $L$.

Finally, using (0.10) and (0.11), the wave function (1.2) equals,
\begin{eqnarray}
\Psi_n(t,z)&=&z^n
e^{\sum_1^{\iy}t_iz^i}\frac{\tau_n(t-[z^{-1}])}{\tau_n(t)}
 \nonumber\\
& =&    \frac{\det \left(z\mu_{ij}(t)-\mu_{i+1,j}(t)
\right)_{0\leq i,j\leq n-1}}
{\det (\mu_{ij}(t))_{0\leq i,j\leq n-1}},~~\mbox{using (0.10), (0.1),}\mbox{ footnote 8,}
 \nonumber\\
&=&\frac{z^n e^{\sum_1^{\iy}t_iz^i}}{n!\det(\mu_{\ell k}(t))}\int
...\int_{\BR^n}\Delta_n(z)
\Delta_n^{(\rho)}(z)\prod^n_{k=1}
\left(e^{\sum
t_iz^i_k}\left(1-\frac{z_k}{z}\right)dz_k\right),\nonumber
\end{eqnarray}
establishing (0.16) and (0.17).\qed

In the subsequent sections, it is shown that many integrable
solutions, when linked together, are nothing but special instances
of the situation described in section 2; we mention matrix
integrals, $2m+1$-band matrices, soliton formulas, the
Calogero-Moser system and others in subsequent sections.

\section{Toda lattice, matrix integrals and
 Riemann-Hilbert for orthogonal polynomials}

\setcounter{equation}{0}

\bigbreak

\medbreak

Setting
\be
\rho_j(u)du:=u^j\rho(u)du,~~~
\rho^t(u):=\rho(u)e^{\sum_1^{\iy}t_ku^k},
\ee
define the moment matrix
\be
m_{\iy}(t):=\left( \mu_{ij}(t)  \right)_{0\leq i,j <\iy}
\mbox{ with}~~\mu_{ij}(t)=\int z^{i+j}e^{\sum_1^{\iy}t_kz^k}\rho(z) dz,\ee and the corresponding $t$-dependent monic orthogonal
polynomials $p_n(t,z)$ in $z$. Note that $m_{\iy}$ is a H\"ankel
matrix and is therefore symmetric. From the form of the moments,
the matrix $m_{\iy}(t)$ satisfies the the following differential
equations
\be
\frac{\pl m_{\iy}}{\pl t}=\Lb^k m_{\iy}.
\ee
Refering to the special case of H\"ankel matrices, discussed at the
end of section 2, we consider the factorization of the symmetric
matrix $m_{\iy}(t)$ into the product of a lower- and
upper-triangular matrix $S_1$ and $S_2$, with $1$'s along the
diagonal of $S_1$ and $h$'s along the diagonal of $S_2$:
\be
m_{\iy}(t)=S_1^{-1}S_2=S_1^{-1}hS_1^{-1 \top}=S^{-1}(t) S^{\top
-1}(t),\mbox{ with }S=h^{-1/2}S_1=h^{1/2}S_2^{-1 \top}.
\ee

\begin{theorem}
Then $S(t)$ and the tridiagonal matrix $L(t)=S(t)\Lb S^{-1}(t)$
satisfy the standard Toda Lattice equations\footnote{with regard to
the splitting of $A \in gl_{\iy}$ into a lower-triangular $A_b$ and
skew-symmetric matrices $A_{sk}$.}:
\be
\frac{\pl S}{\pl t_k}=-\frac{1}{2}(L^k)_b S~~\mbox{and}~~
\frac{\pl L}{\pl t_k}=-\frac{1}{2}[(L^k)_b,L].
\ee
The flag and dual flag of (0.15) take on the following form
\begin{eqnarray}
\WR_n^t&=&\Span\{p_n(t,z),p_{n+1}(t,z),...\}\nonumber\\
&=& \Span\{\left(S(t) \chi(z)\right)_n,\left(S(t)
\chi(z)\right)_{n+1},...\}\nonumber\\
&=&\Span\left\{z^{n}
\frac{\tau_{n}(t-[z^{-1}])}{\tau_n(t) }
,0\leq n <\iy \right\} \nonumber\\ &&\nonumber\\
\WR^{\ast
t}_n &=&
\Span\left\{\int \frac{ p_j (t,u) \rho^t(u)
du}{z-u},j=0,...,n-1\right\}
\oplus
\HR_+\nonumber\\
&=&\Span\left\{
z^{-1}(\left(Sm_{\iy}(t)\right)\chi^{\ast}(z))_j,j=0,...,n-1\right\}
\oplus
\HR_+\nonumber\\
&=&\Span\left\{z^{-j-1}
\frac{\tau_{j+1}(t+[z^{-1}])}{\tau_j(t) }
,j=0,...,n-1\right\}
\oplus
\HR_+,
\end{eqnarray}
with the $\tau$-functions having the following representation,
derived from (0.10) up to (0.13),
\begin{eqnarray}
\tau_n(t)&=&\det \left(\int z^{i+j}\rho^t(z) dz\right)
_{0\leq i,j \leq n-1} \nonumber\\
&=&\frac{1}{n!}\int ...\int_{\BR^n} \Delta^2_n(z)
\prod_{\ell=1}^n
\left(e^{\sum
t_iz^i_{\ell}}\rho(z_{\ell})dz_{\ell}\right)\nonumber\\
 &=& \int_{{\cal
H}_n}e^{Tr (V(M)+\sum_1^{\iy}t_iM^i)}dM,~ \mbox{ setting }
 \rho(z)=e^{V(z)}\nonumber\\
&=&\det
\left(\mbox{Proj}: e^{-\sum_1^{\iy}t_iz^i}z^{-n}\WR_n
\rightarrow
\HR_+\right)\nonumber\\
&=&\left(\int_{\BR}~dz \rho(z)~z^{2(n-1)}  X(t,z)
\right) \tau_{n-1}(t),
\end{eqnarray}
and the orthogonal polynomials, having the form
\begin{eqnarray}
p_n(t,z)&=& z^n \frac{\tau_{n}(t-[z^{-1}])} {\tau_n(t) }
\nonumber\\
 &= &\frac{ \int_{{\cal H}_n}\det (zI-M)e^{Tr
(V(M)+\sum_1^{\iy}t_iM^i)}dM}{\int_{{\cal H}_n}e^{Tr
(V(M)+\sum_1^{\iy}t_iM^i)}dM},\nonumber
\end{eqnarray}
where $dM=\Dt_n^2(z)~dz_1...dz_n~dU$ is Haar measure on the set of
Hermitean matrices ${\cal H}_n$.
\end{theorem}

Before stating the corollary, some explanation is needed. The
integral in the matrix below is taken over the $\BR$
 with a
small upper semi-circle about $z$, when $\Im z >0$ and over $\BR$,
with a small lower semi-circle about z, when $\Im z <0$
. Moreover $Y_{n\pm}(z)=\lim_{\stackrel{z' \rightarrow z}
 {\pm \Im z' >0}} Y_n(z')
$.

\begin{corollary}
In view of the factorization $m_{\iy}(t)=S_1^{-1}S_2$ of the moment
matrix $m_{\iy}(t)$ and setting $h_n=\tau_{n+1}(t)/\tau_n(t)$, we
have the following identity of matrices:
\bea
Y_n(z)&=&
\pmatrix{p_n(t,z)& ~ \int_{\BR}
\frac{p_n(t,u)}{z-u}\rho^t(u)du\cr \cr
h_{n-1}^{-1}p_{n-1}(t,z)& ~h_{n-1}^{-1}\int_{\BR}
\frac{p_{n-1}(t,u)}{z-u}\rho^t(u)du}\nonumber
\\
&&\nonumber
\\
 &=& \pmatrix{\left(S_1 \chi(z) \right)_{n} &
~\frac{1}{z}\left(S_2 \chi^*(z) \right)_{n} \cr \cr
h_{n-1}^{-1}\left(S_1 \chi(z) \right)_{n-1} &
~ h_{n-1}^{-1}\frac{1}{z}\left(S_2 \chi^*(z) \right)_{n-1}
 } \nonumber
\\
&&\nonumber
\\
 &=&
\pmatrix{\frac{\tau_{n}(t-[z^{-1}])} {\tau_n(t) } z^{n}&
\frac{\tau_{n+1}(t+[z^{-1}])}{\tau_n(t)}
z^{-n-1}  \cr
\frac{\tau_{n-1}(t-[z^{-1}])}
{\tau_{n}(t)} z^{n-1}&
\frac{\tau_{n}(t+[z^{-1}])}{\tau_{n}(t) }
z^{-n}  }.
\eea
The matrix $Y_n$ satisfies the Riemann-Hilbert problem of Fokas,
Its and Kitaev
\cite{FIK}:

\noindent 1. $Y(z)$ holomorphic\footnote{$\BC_+$ and $\BC_-$
denote the Siegel upper- and lower half plane.} on $\BC_+$ and $\BC_-$.

\vspace{0.5cm}

\noindent 2. $Y_-(z)=Y_+(z)\pmatrix{1& ~2\pi i\rho^t(z)\cr 0&~ 1}$.

\vspace{0.5cm}
\noindent 3. $Y(z)
=\left( I+O(z^{-1}\right)\pmatrix{z^{n} & 0\cr 0 & z^{-n}}$,
 when $z\rightarrow \iy$.
 \be
 \ee

\noindent Note the first column of $Y(z)$ relates to the Grassmannian
$\WR_n$ and the lower-triangular matrix $S_1$, whereas the second
column to the
 dual $\WR^*_n$ and the upper-triangular matrix $S_2$ in the decomposition of $m_{\iy}=S_1^{-1} S_2$.
\end{corollary}

\underline{\sl Proof of Theorem 4.1}
  The vertex
representation (4.7) of $\tau_n(t)$ shows that the
 $\tau$-vector $\tau(t)=(\tau_n(t))_{n\geq 0}$ is a solution of
 the discrete KP equation (1.4). But more is true: $L=S\Lb S^{-1}$
 is tridiagonal; so, $S$ and $L$ satisfy the {\em standard
 Toda lattice} (4.5). Some of the arguments are contained in
\cite{AvM1}.

Notice that the Borel decomposition (4.4) is tantamount to finding
the orthogonal polynomials $ p_n(t,z)$ with respect to the
inner-product $\la z^i,z^j\ra=\mu_{ij}$, to be precise:
\be
m_{\iy}=S^{-1}S^{\top -1}\Longleftrightarrow Sm_{\iy}S^{\top}=I
\Longleftrightarrow
\la h_i^{-1/2} p_i,
 h_j^{-1/2} p_j \ra =\dt_{ij}.
\ee
It follows that the coefficients of the {\em orthonormal}
polynomials
 $h_i^{-1/2} p_i$ are given by the $i$th row of the matrix
 $S(t)$ and so
 \be
 S_1(t)=h^{1/2}S(t)=(p_{ij}(t))_{0\leq i,j \leq \iy},~\mbox{where} ~
 p_n(t)=\sum_{0\leq j \leq n} p_{nj}(t)z^j.
 \ee

\noindent {\sl (i)} So, the monic polynomials $p_n(t,z)$ of
 (0.17) have the following form:
\begin{eqnarray}
h_n^{1/2} \left( S(t) \chi(z) \right)_n
 &=&
\left( S_1(t) \chi(z) \right)_n\nonumber\\
 &=& p_n(t,z)\nonumber\\
&=&z^n
\frac{\tau_n(t-[z^{-1}])}{\tau_n(t)} \nonumber  \\
&=&\frac{1}{n!\tau_n(t)}\int
...\int_{\BR^n}\Delta^2(u)\prod_{k=1}^n\left((z-u_k)e^{\sum
t_iu^i_k}\rho(u_k)du_k\right),\nonumber  \\
\end{eqnarray}
leading to the formula in the statement of Theorem 4.1.
 The $p_n$'s are the
standard monic {\em orthogonal} polynomials with regard to the
weight $\rho^t(u)=\rho(u)
 e^{\sum t_i u^i}$.

\bigbreak

 \noindent {\sl (ii)} But, we now prove
\bea
 h_n^{1/2} \left( S^{\top -1}(t) \chi^{\ast}(z) \right)_n =
  \left( S_2(t) \chi^{\ast}(z) \right)_n&=&
 z\int \frac{ p_n(t,u)\rho^t(u)}{z-u}du \nonumber\\&=&
z^{-n}\frac{\tau_{n+1}(t+[z^{-1}])}{\tau_n(t) }.
\eea

\noindent Indeed, we compute, on the one hand,
\begin{eqnarray*}
h_n^{1/2}\sum_{j\geq 0} \left(  S m_{\iy} \right)_{nj} z^{-j}
 &=& \sum_{j\geq 0} \left(  S_1 m_{\iy} \right)_{nj} z^{-j}       \\
 &=&\sum_{j\geq 0} z^{-j} \sum_{\ell \geq 0}
 p_{n\ell}(t)\mu_{\ell j},~~\mbox{using (4.11)} \\
 &=&  \sum_{j\geq 0} z^{-j}
\sum_{\ell \geq 0} p_{n\ell}(t)\int_{\BR} u^{\ell+j}\rho^t(u)du
\\ &=&  \int_{\BR} \sum_{\ell \geq 0} p_{n\ell}(t)u^{\ell}
\sum_{j\geq 0}\left(\frac{u}{z}  \right)^j\rho^t(u)du\\
&=&z  \int_{\BR} \frac{ p_n(t,u)\rho^t(u)}{z-u}du.\\
\end{eqnarray*}
On the other hand, as we have seen in the special case following
(2.14), the 2d-Toda $\tau$-function $\tau(t',s')$ depends on
$t=t'-s'$ only, enabling us to write (here $\psi$ stands for $\Psi$
without the exponential),
\begin{eqnarray*}
h_n^{1/2}\sum_{j\geq 0} \left(S^{\top -1}(t)
\right)_{nj} z^{-j} &=&
\left(  h^{1/2} S^{\top -1}(t)\chi(z^{-1})\right)_n  \\
&=& \left(  S_2 (t',s') \chi(z^{-1})\right)_n
 ~,~\mbox{using}~~(4.4),
\\
 &=&
\psi_{2,n}(t',s';z^{-1})~,~\mbox{using}~~(2.5),\\
 &=&\frac{\tau_{n+1}(t',s'-[z^{-1}])}{\tau_n(t',s')
} z^{-n}~,~\mbox{using}~~(2.6),\\
&=&\frac{\tau_{n+1}(t+[z^{-1}])}{\tau_n(t) } z^{-n} ,
\end{eqnarray*}
from which (4.12) follows, upon using $S m_{\iy}=S^{\top -1}$; see (4.9).

Theorem 4.1 is established by remembering Proposition 3.2 and using
(3.8) and (3.9); i.e.,
\begin{eqnarray*}
\WR_n^t&=&\Span\{p_n(t,z),p_{n+1}(t,z),...\}\\
\WR^{\ast t}_n &=&\Span\left\{\int \frac{ u^j
\rho^t(u) du}{z-u},j=0,...,n-1\right\} \oplus \HR_+
\\
&=&\Span\left\{\int \frac{ p_j (t,u) \rho^t(u)
du}{z-u},j=0,...,n-1\right\} \oplus \HR_+,
\end{eqnarray*}
together with (4.12).
\qed

\medbreak

\underline{\sl Proof of Corollary 4.2}:
Following the arguments of Bleher and Its \cite{BI}, the first matrix in (4.8) has the desired
properties taking into account the following integrals:
$$
\frac{1}{2\pi i}\lim_{\stackrel{z' \rightarrow z}{ \Im z' <0}}
\int_{\BR}
\frac{p_n(t,u)}{z'-u}\rho^t(u)du   =
p_n(t,z)\rho^t(z)
  +\frac{1}{2\pi i}\lim_{\stackrel{z' \rightarrow z}{ \Im z' >0}}
  \int_{\BR}
\frac{p_n(t,u)}{z'-u}\rho^t(u)du.
$$
The formulas (4.11) and (4.12) lead to the desired result.
\qed

\remark From the fact that $\det Y_{n-}=\det Y_{n+}$, it follows
 that $\det Y(z)$ is holomorphic in $\BC$ and since $\det Y(z)
 =1+O(z^{-1})$, it follows from Liouville's theorem that
 $\det Y(z)=1$, i.e.,
 \bea
\det Y_n &= & h_{n-1}^{-1} \left(p_n(t,z) \int_{\BR}
\frac{p_{n-1}(t,u)}{z-u}\rho^t(u)du
- p_{n-1}(t,z) \int_{\BR}
\frac{p_{n}(t,u)}{z-u}\rho^t(u)du   \right)\nonumber\\
&=&\frac{1}{\tau_n^2(t)}
\left(\tau_{n}(t-[z^{-1}])\tau_{n}(t+[z^{-1}])
 -z^{-2}\tau_{n-1}(t-[z^{-1}])\tau_{n+1}(t+[z^{-1}])
 \right) \nonumber\\ &=&1.
 \eea
 This is not surprising, in view of the fact that the
 first expression for $\det Y_n$ is nothing but the Wronskian
 of the the two fundamental solutions of the second
 order difference equation; see Akhiezer \cite{A}.
 The second expression, involving $\tau$-functions follows also from Corollary 2.2, by setting
 $u=z^{-1}$ and $v\rightarrow 0$ and by using the fact that, for
 the standard Toda lattice, we have $\tau(t,s)=\tau(t-s)$.


\section{\bf Periodic sequences of weights,
$2m+1$-band matrices and Riemann-Hilbert problems}

The results of section 4 about tridiagonal matrices will be
extended in this section to $2m+1$-band matrices. As usual, we
 set
\be
\mu_{ij}(t)=\left\la z^i, \rho_j^t(z)\right\ra, ~\mbox{with}~~
\rho_j^t(z)=e^{\sum t_k z^k} \rho_j(z).
\ee
In proving and stating the results below, we shall also consider the
$s$-deformations, as in (2.1). Here we consider $m$-periodic
sequences of weights $\rho_0,\rho_1,...$, defined by
\be
\rho_{j+km}(z)=z^{km}\rho_j(z),\quad\mbox{for all
$j=0,1,2,...$}.
\ee
\medbreak

\begin{theorem} For the weights (5.2), the polynomials
\bea
p_n(t,z)&=&
\frac{1}{\det\left(\mu_{\ell,k}(t)\right)_{0\leq\ell,k\leq
n-1}}\det\left(\begin{array}{ccc|c}
\mu_{00}(t)&...&\mu_{0,n-1}(t)&1\\
\vdots& &\vdots&\vdots\\
\mu_{n-1,0}(t)&...&\mu_{n-1,n-1}(t)&z^{n-1}\\ \hline
\mu_{n0}(t)&...&\mu_{n,n-1}(t)&z^n
\end{array}\right)\nonumber\\
 &=&z^n
\frac{\tau_n(t-[z^{-1}],0)}{\tau_n(t,0)},~\mbox{where}~~
 \tau_n(t,0)=det~ m_n(t)\nonumber\\
&=&    \frac{\det \left(z\mu_{ij}(t)-\mu_{i+1,j}(t)
\right)_{0\leq i,j\leq n-1}}
{\det (\mu_{ij}(t))_{0\leq i,j\leq n-1}}\nonumber \\
&=&\frac{1}{n!\tau_n(t)}
\int ...\int_{\BR^n}\Delta_n(z)
\Delta_n^{(\rho)}(z)\prod^n_{k=1}
\left(e^{\sum_i
t_iz^i_k}\left(z-z_k \right)dz_k\right),
\eea
lead to matrices $L$, defined by $zp=Lp$,

\bigbreak

\noindent $
\left\{
\begin{tabular}{l}
(i)~ which evolve according to the discrete KP hierarchy
\\ (ii)~such that $L^m$ is a
\underline{$2m+1$-band matrix}.\\
(iii)~the polynomials $p_n(z)$ satisfy the {\em generalized
orthogonality relations}
 \\ ~~~~~~ $\la p_i(z),
\rho_j(z) \ra=0$ for $i \geq j+1$.      \\
\end{tabular}
\right.
$

\end{theorem}

\vspace{0.5cm}

\noindent \remark It is interesting to point out that the condition (5.2) is
equivalent to a seemingly weaker one:
\be
z^m\rho_j\in\Span\{\rho_0,...,\rho_{m+j}\},\quad\mbox{for all
$j=0,1,2,...$}, \ee where $\rho_{m+j}$ must appear in the span.
Indeed, the $p_n$'s only depend on the moments $\mu_{ij}$ by means
of the determinantal formulae (5.3), which allow for column
operations.

\begin{corollary}
The following $2\times 2$ matrices are all equal
\bea
Y_n(z)&=& \pmatrix{p_n(t,z)&~~~ \displaystyle{
\int_{\BR}
\frac{p_n(t,u)}{z^m-u^m}\left(\sum_{k=1}^m
z^{m-k}\rho_{k-1}^t(u)\right)du}\cr h_{n-1}^{-1}
 p_{n-1}(t,z)&~~~
h_{n-1}^{-1}\displaystyle{\int_{\BR}
\frac{p_{n-1}(t,u)}{z^m-u^m}\left(\sum_{k=1}^m
z^{m-k}\rho_{k-1}^t(u)\right)du}    }\nonumber\\ &&
\nonumber\\
&&\nonumber
\\
 &=& \pmatrix{\left(S_1 \chi(z) \right)_{n} &
\frac{1}{z}\left(S_2 \chi^*(z) \right)_{n} \cr
\cr
h_{n-1}^{-1}\left(S_1 \chi(z) \right)_{n-1} &
 h_{n-1}^{-1}\frac{1}{z}\left(S_2 \chi^*(z) \right)_{n-1}
 } \nonumber
\\
&&\nonumber
\\
&=&\pmatrix{ \displaystyle{\frac{\tau_{n}(t-[z^{-1}] ,0)}
{\tau_n(t,0) } z^{n}}& \displaystyle{
\frac{\tau_{n+1}(t,-[z^{-1}])}{\tau_n(t,0)}z^{-n-1}}
  \cr
\displaystyle{\frac{\tau_{n-1}(t-[z^{-1}],0)}
{\tau_{n}(t,0)}z^{n-1}} &
\displaystyle{\frac{\tau_{n}(t,-[z^{-1}])}
{\tau_{n}(t,0) }
 z^{-n} } }; \nonumber \\
\eea
they solve the following Riemann-Hilbert problem:

\noindent 1. $Y_n(z)$ holomorphic on $\BC_+$ and $\BC_-$.

\vspace{0.5cm}

\noindent 2. $Y_{n-}(z)=Y_{n+}(z)\pmatrix{1& \frac{2\pi i}
{m} e^{\sum t_k z^k}
\sum_{1}^{m}z^{-j} \rho_j(z)\cr 0& 1}$.

\vspace{0.5cm}
\noindent 3. $Y_n(z)\pmatrix{z^{-n} & 0\cr 0 & z^n}
\longrightarrow I$, as $z\rightarrow \iy$.

\vspace{0.5cm}
\noindent 4. $Y_n(z)\pmatrix{1 & 0\cr 0 & z^{m-1}}$ finite
, as $z\rightarrow 0$.

\noindent The
polynomials $p_n$ are such that $z^m p_n$ satisfies
\underline{$2m+1$-step relations}.

Note
$$
\det Y_n=\frac{\tau_n(t-[z^{-1}],-[z^{-1}])}{\tau_n(t,0)}
$$
and the first column of  $Y_n$ is related
 to the Grassmannian plane
$\WR_n$ and the second column to a plane\footnote{Unlike the
orthogonal polynomial case, the second column does not contain
elements of the dual Grassmannian $\WR^*$.} related to
$\Psi_2(t,s,z)$.

\end{corollary}

\remark In the matrix (5.5), $\tau_n(t-[z^{-1}],0)$ is given in
formula (5.3), whereas by (0.10) and (5.11),
\be
\tau_{n}(t,-[z^{-1}])=\det \left(\left\la u^i,
e^{\sum_1^{\iy}t_i u^i} \sum_{k=0}^{m-1}
  \frac{z^{m-k}\rho_{k+j}(u)}{z^m-u^m} \right\ra
  \right)_{0\leq i,j \leq n-1}.
\ee Also note the right hand column of (5.5) behaves as
 $1/z^{n+1}$; this follows from the $\tau$-function
 representation, but also from the "generalized
 orthogonality", mentioned in (iii) (Theorem 5.1).

Proving Theorem 5.1 requires the following lemma:

\medbreak

\begin{lemma} Fix $m\geq 1$; the polynomials $p_n(t,z)$, defined
in (0.2), satisfy $2m+1$-step recursion relations, i.e., $$
z^mp(t,z)=L^mp(t,z)\mbox{\,\,with\,\,$L^m$ a $2m+1$-band matrix,}
$$ if and only if every $\rho_j$, $j=0,1,...$ satisfies the
following requirement: $$ \left\{\begin{tabular}{l} For every
$\ell=0,1,...$, $j=0,1,...$ there exist constants $c_r$,
$r=0,...,m+j+\ell$\\ depending on $j$ and $\ell$ such that\\
\hspace{1cm} $\la u^m\rho_j-\sum_0^{m+j+\ell}c_r\rho_r,u^i\ra =0$
for
 $0\leq i\leq m+j+\ell+1$
\end{tabular}
\right.
$$
\end{lemma}

\medbreak

\proof  Note the following
equivalences:
$$
\begin{tabular}{lll}
 &$z^mp_n(t,z)=\displaystyle{\sum_{n-m\leq r\leq
n+m}}A(t)_{nr}p_r(t,z)$&for some matrix $A(t)$\\ & & \\
 &$\Longleftrightarrow$~$z^mp_n(t,z)\in \WR^t_{\max(n-m,0)}$&for all $n\geq 0$\\
& & \\
 &$\Longleftrightarrow$~$z^m\WR_n^t\subset \WR^t_{\max(n-m,0)}$&for all $n\geq 0$, because
\\
 & & of the inclusion \\
 & & $...\supset \WR_n\supset \WR_{n+1}\supset ...,$\\
& & \\
 &$\Longleftrightarrow$~$z^m\WR_n\subset \WR_{\max(n-m,0)}.$&
\end{tabular}
$$
Since
$$
\WR_n=(\Span\{\rho_0,\rho_1,...,\rho_{n-1}\})^{\bot}=\Span\{p_n(z),p_{n+1}
(z),...\}$$
the latter is equivalent to
\begin{eqnarray*}
0&=&\la u^mp_n(u),\rho_j(u)\ra~~\mbox{for all $0\leq j\leq n-m-1$,
 $n\geq 0$}\\ &=&\la p_n(u),u^m\rho_j(u)\ra\\
&=&\frac{1}{a_{nn}(t)}\sum^n_{i=0}a_{ni}\la u^i,u^m\rho_j(u)\ra\\
&=&\frac{1}{a_{nn}(t)}\det\left(\begin{tabular}{lll|l}
$\mu_{00}$&...&$\mu_{0,n-1}$&$\mu_{mj}$\\ $\vdots$&
&$\vdots$&$\vdots$\\
\hline
$\mu_{n0}$&...&$\mu_{n,n-1}$&$\mu_{m+n,j}$
\end{tabular}\right),
\end{eqnarray*}
where we have used the fact that $p_n(t,z)=\frac{1}{a_{nn}(t)}\sum
a_{ni}(t) z^i$ is represented by (0.2). The vanishing of the
determinant above is equivalent to the statement that the last
column depends on prior columns; namely there exist
$c_0,...,c_{n-1}$ depending on $m,n,j$ such that
\begin{eqnarray*}
0&=&\mu_{m+i,j}-\sum_{r=0}^{n-1}c_r \mu_{ir}\quad\mbox{for
$0\leq i\leq n$, $0\leq j\leq n-m-1$}\\ &=&\la u^{i+m},\rho_j\ra
-\sum_{r=0}^{n-1}c_r \la u^i,\rho_r\ra\\
&=&\la u^i,u^m\rho_j-\sum_{r=0}^{n-1}c_r \rho_r\ra\\ &=&\la
u^i,u^m\rho_j-\sum_{r=0}^{j+m+\ell}c_r\rho_r\ra\quad\mbox{for $
0\leq i \leq m+j+\ell+1$},
\end{eqnarray*}
where $\ell$ was defined such that $j+\ell=n-m-1$.
\qed

\underline{\sl Proof of Theorem 5.1}: The fact that
 $L^m$ is a $2m+1$-band matrix follows at once from
 Lemma 5.3. That the matrix $L$ evolves according to
 the discrete KP-hierarchy follows straightforwardly
 from the general statement in Theorem 0.1.\qed

\underline{\sl Proof of Corollary 5.2}:
 For the sake of this proof, we shall be using the
 $(t,s)$-deformations of the weights $\rho_j$ and the
 corresponding matrix $m_{\iy}(t,s)$ of
 $(t,s)$-dependent moments
 \be
\mu_{ij}(t,s)=\left\la z^i, \rho_j(t,s;z)\right\ra,
 ~\mbox{with}~~
\rho_j(t,s;z)=e^{\sum_1^{\iy}t_i z^i}
 \sum_{\ell=0}^{\iy} F_{\ell}(-s)\rho_{j+\ell}(z)
.
\ee
In section 2, it was mentioned that $m_{\iy}$
 satisfies the differential equations
\be
\frac{\pl m_{\iy}}{\pl t_k}=\Lb^k m_{\iy}~,~~~ \frac{\pl
m_{\iy}}{\pl s_k}=-m_{\iy} \Lb^{k \top} .\ee Factorizing the
matrix
\be
m_{\iy}(t,s)= S_1^{-1}(t,s) S_2(t,s)
\ee
 into the product of lower- and
upper-triangular matrices $S_1$ and $S_2$ then leads to the 2d Toda
lattice.

For later use, we also compute $\rho_k(t,-[z^{-1}];u)$, making
specific use of the periodicity of the sequence of weights
$\rho_{j+m}(u)=u^m
\rho_j(u)$
and the identity\footnote{obtained, by expanding the following
expression in elementary Schur polynomials, by setting $t=0$ and by
comparing the powers of $y$:
 $$
\sum_{n\geq 0} y^n F_n(t+[z^{-1}])=
e^{\sum \left( t_i+\frac{z^{-i}}{i}  \right)y^i}=
 e^{\sum  t_i y^i}\left( 1-\frac{y}{z} \right)^{-1}
 = \sum_{n\geq 0} y^n F_n(t)\sum_0^{\iy}
 \left( \frac{y}{z}\right)^k.$$
 }
\be
F_n([z^{-1}])=z^{-n}.
\ee
 We find:
\begin{eqnarray}
\rho_k(t,-[z^{-1}];u)&=&e^{\sum_1^{\iy}t_i u^i}
 \sum_{j=0}^{\iy} F_{j}([z^{-1}])\rho_{j+k}(u)\nonumber\\
 &=& e^{\sum_1^{\iy}t_i u^i}\sum_{j=0}^{\iy}
  \frac{\rho_{j+k}(u)}{z^j},~~\mbox{using again (),}\nonumber\\
&=& e^{\sum_1^{\iy}t_i u^i}\left(\sum_{j=0}^{m-1}
  \frac{\rho_{j+k}(u)}{z^j} +
  \sum_{j=m}^{2m-1}
  \frac{\rho_{j+k}(u)}{z^j}+...  \right)\nonumber\\
&=& e^{\sum_1^{\iy}t_i u^i} \sum_{j=0}^{m-1}
  \frac{\rho_{j+k}(u)}{z^j}\left(1 +
 (\frac{u}{z})^m +
  (\frac{u}{z})^{2m}+...\right)\nonumber\\
&=& e^{\sum_1^{\iy}t_i u^i} \sum_{j=0}^{m-1}
  \frac{\rho_{j+k}(u)}{z^j(1-(\frac{u}{z})^m)}   \nonumber \\
&=&  e^{\sum_1^{\iy}t_i u^i} \sum_{j=0}^{m-1}
  \frac{z^{m-j}\rho_{j+k}(u)}{z^m-u^m}.
\end{eqnarray}

 From (5.9), we have $S_1 m_{\iy} =S_2$ and hitting $\chi^*(z)$
 with this
 matrix, we compute, on the one
 hand,
\begin{eqnarray}
\sum_{j\geq 0} \left(  S_1 m_{\iy} \right)_{nj}
 z^{-j}\Big|_{s=0}
 &=&\sum_{j\geq 0} z^{-j} \sum_{\ell \geq 0}
 p_{n\ell}(t,s)\mu_{\ell j} \Big|_{s=0}\nonumber \\
&=&\sum_{j\geq 0} z^{-j} \sum_{\ell \geq 0}
 p_{n\ell}(t,s)\int_{\BR}u^{\ell}\rho_j(t,s;u)du
   \Big|_{s=0}\nonumber\\
 &=&  \sum_{j\geq 0} z^{-j}\int_{\BR}p_n(t,s;u)
 e^{\sum_1^{\iy}t_i u^i}
\sum_{\ell \geq 0} F_{\ell}(-s)\rho_{j+\ell}(u)du
 \Big|_{s=0} \nonumber\\
&=&  \int_{\BR}  p_{n}(t,0;u)e^{\sum_1^{\iy}t_i u^i}
\sum_{j\geq 0}  z^{-j}   \rho_j(u)du\nonumber\\
&=&    \int_{\BR}  p_{n}(t,0;u)e^{\sum_1^{\iy}t_i u^i}
\sum_{j\geq 0}  F_j([z^{-1}])   \rho_j(u)du \nonumber  \\
&=&    \int_{\BR}  p_{n}(t,0;u)
\rho_0(x,-[z^{-1}];u)du \nonumber\\
 &=&    \int_{\BR}  p_{n}(t,0;u)
e^{\sum_1^{\iy}t_i u^i} \sum_{j=0}^{m-1}
  \frac{z^{m-j}\rho_{j}(u)}{z^m-u^m}du,
\end{eqnarray}
using (5.11) in the last identity. On the other hand, we have
\begin{eqnarray}
\sum_{j\geq 0} \left(S_2(t,s)
\right)_{nj} z^{-j}\Big|_{s=0} &=&
\left(  S_2(t,s)\chi(z^{-1})\right)_n \Big|_{s=0}
 \nonumber\\
 &=&
\psi_{2,n}(t,0;z^{-1})\nonumber\\
 &=&\frac{\tau_{n+1}(t,-[z^{-1}])}{\tau_n(t,0)
} z^{-n}.
\end{eqnarray}

The right hand sides of (5.12) and (5.13) coincide (using
$S_1m_{\iy}=S_2$); so, we find the following identity, together
with the desired
 asymptotics for $z \rightarrow \iy$:
\be
 \int_{\BR}  \frac{p_n(t;u)}{z^m-u^m}
 \left(\sum_{j=0}^{m-1}z^{m-j}
\rho_j^t(u)\right)du
=z^{-n}\frac{\tau_{n+1}(t,-[z^{-1}])}{\tau_n(t,0)
}=z^{-n}(h_{n}+O(1)).
\ee
leading to condition 3. The jump condition 2. follows from the
following
$$
\frac{1}{2\pi i} \lim_{\stackrel{z' \rightarrow z}
 {\Im z' <0}}
\int
\frac{p_n(t,u)}{z^m-u^m}\left(\sum_{j=0}^{m-1}z^{m-j}
\rho_j^t(u)\right)du \hspace{5cm}
$$
\be
 = p_n(t,z)\frac{1}{m}\sum_{j=0}^{m-1}z^{1-j}\rho_j^t(z)
  +\frac{1}{2\pi i}\lim_{\stackrel{z' \rightarrow z}
 { \Im z' >0}}\int
\frac{p_n(t,u)}{z^m-u^m}\left(\sum_{j=0}^{m-1}z^{m-j}
\rho_j^t(u)\right)du.
\ee
The asymptotics 3. follows from the $\tau$-function representation
of the first integral.

The formula concerning $\det Y_n$ follows from setting $u=0$ and
$v=z^{-1}$ in identity (2.8).
\qed

\vspace{0.6cm}

\section{\bf Soliton formula}

\medbreak

For future use, we define the vertex operator:
\be
X(t;\lb,\mu)=\frac{1}{\lb-\mu}e^{\sum_1^n (\lb^k-\mu^k)t_k}
e^{\sum^n_{1}(\mu^{-k}-\lb^{-k})\frac{1}{k}\frac{\pl}{\pl t_k}}.
\ee

\begin{theorem}
 Given points $p_k,q_k$ and $\lb_k$, $k=1,...$, and the weights
$$
\rho_k=\delta(z-p_{k+1})-\lb^2_{k+1}\delta(z-q_{k+1}),\quad
k=0,1,...\,,
$$
the $\tau$-functions
\begin{eqnarray*}
\tau_n(t)&=&\det\left(p_i^{j-1}e^{\sum_{k=1}^{\iy}t_kp_i^k}-\lb^2_iq_i^{j-1}
e^{\sum_{k=1}^{\iy}t_kq_i^k}\right)_{1\leq i,j\leq n}\\
 &=& \left(
p_n^{n-1}X(t,p_n)-\lb^2_n q_n^{n-1}X(t,q_n)\right) \tau_{n-1}(t)\\
&=&c_n
\prod_{k=1}^ne^{\sum t_i p_k^i}\,\det\left(\delta_{ij}-\frac{a_i}{q_i-p_j}
e^{\sum_{k=1}^{\iy}t_k(q_i^k-p_j^k)}\right)_{1\leq i,j\leq n}\\
&=&c_n
\prod_{k=1}^ne^{\sum t_i p_k^i}\prod^n_1e^{-a_iX(t;q_i,p_i)}1,
\end{eqnarray*}
form a $\tau$-vector of the discrete KP hierarchy, for
appropriately chosen functions $a_i$ and $c_n$ of $p,q,
\lb$. The matrix $L$, constructed by (0.9) from the $\tau$'s above,
satisfies $$L(t)p(t,z)=zp(t,z),$$ with polynomial eigenvectors (in
$z$):
$$
p_n(z)=\frac{\det\left(\delta_{ij}(z-p_i)-\frac{a_i(z-q_i)}{q_i-p_j}
e^{\sum_{k=1}^{\iy}t_k(q_i^k-p_j^k)}\right)_{1\leq i,j\leq
n}}{\det\left(\delta_{ij}-\frac{a_i}{q_i-p_j}
e^{\sum_{k=1}^{\iy}t_k(q_i^k-p_j^k)}\right)_{1\leq i,j\leq n}}.
$$
Then
\bea
\WR_n &=&(\Span\{\rho_0,...,\rho_{n-1}\})^{\bot}\nonumber\\
&=&\{f\in\HR^+,\mbox{\,\,such that
$f(p_i)=\lb^2_if(q_i),i=1,...,n\}$} \\
\mbox{and}\hspace{1.5cm}  & & \nonumber\\
\WR^*_n
&=&\Span\left\{\frac{1}{z-p_i}-\frac{\lb^2_i}{z-q_i},i=1,...,n\right\}\oplus
\HR_+.
\eea
\end{theorem}

\proof Consider the space
\begin{eqnarray}
\HR^+/z^n\HR^+&=&\Span\{1,...,z^{n-1}\}\nonumber\\
&=&\Span\{v_1,...,v_n\},
\end{eqnarray}
where the polynomials
$$
 v_k(z)=\prod_{{j\neq k}\atop{1\leq j\leq
n}}(z-p_j)
 $$
 of degree $n-1$ form an alternative basis;
 the determinant of the
 transformation between the two bases being the
 Vandermonde determinant $\Delta_n(p)=\prod_{1\leq
 j<k \leq n}(p_k-p_j)$. Define
\be
P(z):=\prod^n_1 (z-p_i)
\ee
and
\begin{eqnarray*}
g(z)=e^{\sum_1^{\iy}t_iz^i},~~~a_k=\lb^2_k\frac{P(q_k)}{v_k(p_k)},~~~
c_n(t)&=&\frac{\prod_1^n
\left(P'(p_k)e^{\sum_{i=1}^{\iy}t_ip_k^i}\right)} {\Dt_n(p)}
\\ &=&c_n
\prod_{k=1}^n e^{\sum_{i=1}^{\iy}t_ip_k^i}.
\end{eqnarray*}
With this notation
\be
v_k(p_i)=\delta_{ik}v_k(p_k)=\dt_{ik}P'(p_k), ~~~~~
v_k(q_i)=\frac{P(q_i)}{q_i-p_k}.
\ee
Using the second line of (3.14) and the formula for $\rho_i$, one
computes
\begin{eqnarray*}
\tau_n(t)&=&\det(
\mu_{ji})_{0\leq i,j\leq n-1}\\
&=&\det\left(\oint_{z=\iy}z^{-j-1}g(z)z^n\left(\frac{1}{z-p_i}-
\frac{\lb^2_i}{z-q_i}\right)\frac{dz}{2 \pi i}\right)_{\,_{{1\leq i\leq
n}\atop{0
\leq j\leq n-1}}}\\
&=&\frac{1}{\Delta(p)}\det\left(\oint_{z=\iy}v_j(z)g(z)\left(\frac{1}{z-p_i}-
\frac{\lb^2_i}
{z-q_i}\right)\frac{dz}{2 \pi i}\right)_{1\leq i,j\leq n},
\end{eqnarray*}
using (6.4). The second identity above leads to the first formula
in the statement above about $\tau$, whereas (0.14)
 is responsible for the second formula. The last
 formula on the right hand side,just above, leads to
\begin{eqnarray*}
\tau_n(t)&=&\frac{1}{\Delta
(p)}\det\left(v_j(p_i)g(p_i)-v_j(q_i)g(q_i)\lb^2_i\right)_{1\leq
i,j\leq n}\\ &=&\frac{1}{\Delta (p)}\det
\Bigl[\diag(g(p_i))_{1\leq i\leq n}\\ & &
\hspace{1.2cm}\left(\delta_{ij}v_i(p_i)g(p_i)-v_j(q_i)g(q_i)\lb^2_i\right)_{1\leq
i,j\leq n} \diag\left(g(p_i)^{-1}\right)_{1\leq i\leq n}\Bigr]\\
&=&\frac{1}{\Delta (p)}
\det\left(\delta_{ij}v_i(p_i)g(p_i)-\lb^2_iv_j
(q_i)g(q_i)\frac{g(p_i)}{g(p_j)}\right)_{1\leq i,j\leq
n},~\mbox{using}~(6.4)\\
&=&c_n(t)\,\det\left(\delta_{ij}-\lb^2_i\frac{v_j(q_i)}{v_i(p_i)}
\frac{g(q_i)}{g(p_j)}\right)_{1\leq i,j\leq n}\\
&=&c_n(t)\,\det\left(\delta_{ij}-\frac{a_i}{q_i-p_j}
\frac{g(q_i)}{g(p_j)}\right)_{1\leq i,j\leq n}\\
&=&c_n(t)\,\det\left(\delta_{ij}-a_iX(t,q_i,p_j)1\right)_{1\leq
i,j\leq n}\\ &=&c_n(t)\,\prod^n_{i=1}e^{-a_iX(t,q_i,p_i)}1,
\end{eqnarray*}
using in the last equality the vanishing of the square of the
vertex operator.

The formula for $p_n(t,z)$ is derived from the third expression for
$\tau(t)$, using the standard representation (2.6) for the wave
vector $\Psi(t,z)$. \qed
\medbreak
\underline{Remark}: When $q_i=-p_i$, the formula for the KdV
$\tau$-function reads:
\begin{eqnarray*}
\tau_n (t)&=&\left(\prod^n_1 p_i\right)^n\prod^n_1\lb_i
e^{\sum_{k,i}t_{2k}p_i^{2k}}\\ & &
\hspace{1cm}\det\left(p_i^{-j}\left(\lb^{-1}_i e^{\sum_{{\rm
odd}}t_kp_i^k}-(-1)^{n-j}\lb_ie^{-\sum_{{\rm
odd}}t_kp^k_i}\right)\right)_{1\leq i,j\leq n}\\
&=&c_n(t)\,\det\left(\delta_{ij}+\frac{a_i}{p_i+p_j}e^{-\sum_{{\rm
odd}}t_k(p_i ^k+p_j^k)}\right)_{1\leq i,j\leq n}.
\end{eqnarray*}
Note that Segal and Wilson have used, in \cite{SW},
 the infinite matrix representation of the projection
 of (6.2), rather than (6.3), in order to compute
 KdV solitons.

\bigbreak

\section{\bf Calogero-Moser system}

\medbreak

 \begin{theorem} Given points $p_k,\lb_k$ ($k=1,2,...$), the weights
$$
\rho_k=\dt'(z-p_{k+1})+\lb_{k+1}\dt(z-p_{k+1}),\quad k=0,1,...,n-1
$$
determine a sequence of $\tau$-functions for the discrete KP
equation\footnote{$\bar t=(x,t_2,t_3,...)$.},
\begin{eqnarray}
\tau_n(t)&=&\frac{1}{n!}\int
...\int_{\BR^n}\prod_{k=1}^n
e^{\sum t_i z^i_k}
\Delta_n(z)\Delta_n^{(\rho)}(z)dz_1...dz_n\nonumber\\ &=&e^{\tr
\sum^{\iy}_1 t_i Y^i}\,\det\left(-X+\sum_1^{\iy}k\bar t_kY^{k-1}
\right),
\end{eqnarray}
with appropriate matrices $X$ and $Y$, functions of $p_k$ and
$\lb_k$'s, satisfying the commutation relation\footnote{where
$I_e=(1-\dt_{ij})_{1\leq i,j\leq n}$.} $[X,Y]=I_e$, and having the
form
\be
 X=\diag(x_1,...,x_n)~~\mbox{and}~~
Y=\left(\frac{1-\dt_{ij}}{x_i-x_j}\right)_{ij}+\diag(\xi_1,...,\xi_n).
\ee
The matrix $L$, constructed by (0.9) from the $\tau$'s above,
satisfies
$$
L(t)p(t,z)=z p(t,z),
$$
with eigenvectors, polynomial in $z$,
\be
p_n(\bar t,z)= \det\left(zI-Y-
\left(xI+\sum_1^{\iy}kt_kY^{k-1} - X \right)^{-1}
\right)
.\ee
The Grassmannian flag corresponding to this construction is given
by
\begin{eqnarray}
\WR_n&=&\{f\in \HR^+,\mbox{\,\,such that $f'(p_i)=\lb_i\,f(p_i),1\leq
i\leq n\}$},\nonumber\\
\WR^*_n&=&\left\{\frac{1}{(z-p_i)^2}-\frac{\lb_i}{z-p_i},i=1,...,n\right\}
\oplus \HR_+.
\end{eqnarray}
\end{theorem}

\proof As before, we introduce the basis $v_k(z)$ of $\HR^+/z^n \HR^+$;
note
\be
\frac{\pl}{\pl z}v_k=\sum^n_{i=1}\prod_{j\neq k,i}(z-p_j)=\sum_{{i=1}
\atop{i\neq k}}\frac{v_k(z)-v_i(z)}{p_k-p_i}.
\ee
and the matrices
\bea
\tilde X&=&-\diag\left(\sum_{{1\leq\al\leq n}\atop{\al\neq
i}}\frac{1}{p_i-p_{\al}}-\lb_i\right)_{1\leq i\leq n}-\left(
\frac{1-\dt_{ij}}{p_i-p_j}\right)_{1\leq i,j\leq n}\nonumber\\
\tilde Y&=&\diag(p_1,...,p_n)
\eea
with commutation relation
\be
[\tilde X,\tilde Y]=I_e,\quad I_e=(1-\dt_{ij})_{1\leq i,j\leq n}.
\ee
Then, by (3.7) and the choice\footnote{$c_n:=\frac{\prod_1^n
v_i(p_i)}{\Dt_n(p)}$in the expressions below.} of $\rho_i$,

\bigbreak

\noindent$\tau_n(t)$
\begin{eqnarray*}
&=&\det(\mu_{ij})_{0\leq i,j\leq n-1}\\ &=&\det\left(\oint_{z=\iy}
z^{j-1}g(z)\left(\frac{1}{(z-p_i)^2}-
\frac{\lb_i}{z-p_i}\right)\frac{dz}{2 \pi i}\right)_{1\leq i,j\leq n}\\
&=&\frac{1}{\Delta_n(p)}\det\left(\oint
v_j(z)g(z)\left(\frac{1}{(z-p_i)^2}-
\frac{\lb_i}{z-p_i}\right)\frac{dz}{2 \pi i}\right)_{1\leq i,j\leq n}\\
&=&\frac{1}{\Delta_n(p)}\det\left((v_j
g)^{\prime}\Bigl|_{z=p_i}-\lb_iv_j(p_i)g(p_i)\right)_{1\leq i,j\leq
n}\\
&=&\frac{1}{\Dt_n(p)}\det\left(g(p_i)\sum^n_{{\al=1}\atop{\al\neq
j}}\frac{v_j(p_i)-v_{\al}(p_i)}{p_j-p_{\al}}+v_j(p_i)g(p_i)
\left(\sum^{\iy}_1kt_kp_i^{k-1}-\lb_i\right)\right)_{1\leq
i,j\leq n}\\ &=&c_n\prod_{k=1}^n e^{\sum_1^{\iy}t_i
p_k^i}\,\det\left(\frac{1-\dt_{ij}}{p_i-p_j}+\dt_{ij}\left(
\sum^n_{{\al=1}\atop{\al\neq
i}}\frac{1}{p_i-p_{\al}}+\sum^{\iy}_1kt_kp_i^{k-1}-\lb_i\right)
\right)_{1\leq i,j\leq n}\\
&=&c_ne^{\tr
\sum^{\iy}_1 t_i \tilde Y^i}\,\det\left(-\tilde X+\sum_1^{\iy}kt_k\tilde Y^{k-1}\right),
\end{eqnarray*}
yielding the formula for $\tau_n(t)$; According to theorem 0.1, the
$p_n(t,z)$ are polynomials, which we now compute:
\begin{eqnarray}
p_n(t,z)&=&z^n \frac{\tau_n(\bar t-[z^{-1}])}{\tau_n(\bar
t)}\nonumber\\ &=&z^n\prod_1^n
\left( 1-\frac{p_k}{z}
\right)\frac{\det\left(-\tilde X+
\displaystyle{\sum^{\iy}_1}k\bar t_k
\tilde Y^{k-1}-z^{-1}
\displaystyle{\sum_1^{\iy}}
\left(\frac{\tilde Y}{z}\right)^{k-1}\right)}
{\det\displaystyle{\left(-\tilde X+\sum_1^{\iy}k\bar t_k
Y^{k-1}\right)}}\nonumber\\ & & \nonumber\\
&=&\det(zI-Y)~\frac{\det\displaystyle{\left(-\tilde X+\sum_1^{\iy}
k\bar t_k
\tilde Y^{k-1}-z^{-1}\left(1-z^{-1}\tilde Y\right)^{-1}\right)}}{\det\displaystyle{\left(-\tilde
X+
\sum_1^{\iy}k\bar t_k \tilde Y^{k-1}\right)}}\nonumber\\
& & \nonumber\\ &=&\det(zI-\tilde
Y)~\det\left(I-\left(xI+\sum_1^{\iy}kt_kY^{k-1}-\tilde
X\right)^{-1} (z-\tilde Y)^{-1}\right),\nonumber\\
\end{eqnarray}
yielding (7.3), but also an expression for the wave functions
$\Psi_n(t,z)$, upon multiplying by an exponential. The formulae for
$\WR_n$ and $\WR^{\ast}_n$ follow from (3.8) and (3.9) and the
choice of $\rho_k$.

In order to connect with the form of the matrices announced in
(7.2), consider the hyperplane $V$ perpendicular to
$e=(1,...,1)\in\BC^n$
$$
\BC^n\supset V=\{\langle z,e\rangle =0\}
$$
and the isotropy subgroup $G_e\in U(N)$ of $I_{e}$, i.e., the $U$'s
such that $U^{\top}e=e$, thus preserving $V$. That $I_e=-I\Bigl|_V$
follows at once, from
$$
I_ez=\left(\sum_{k\neq i}z_k\right)_{1\leq i\leq n}=-z.
$$
Since $G_e$ stabilizes $I_{e}$, there exists a unitary matrix $U\in
G_e$, diagonalizing $X$, having the property
$$
[U\tilde X U^{-1},U\tilde Y U^{-1}]=[X,Y]=UI_eU^{-1}=I_e,
$$
with
$$
X=\diag(x_1,...,x_n);
$$
i.e.,
$$
(x_i-x_j) y_{ij}=1-\dt_{ij}\quad \quad i\neq j,
$$
implying $ Y$ must have the form announced in (7.2).
 Introducing
these new matrices into the expressions for $\tau_n$ and $p_n(t,z)$
yields
\begin{eqnarray*}
\tau_n(t)&=&e^{\tr \sum_1^{\iy} t_i  Y^i}\det\left( -
X+\sum_1^{\iy}kt_k Y^{k-1}\right)\\
 &=&\det e^{ \sum_1^{\iy} t_i  Y^i}\det(-
X+t_1I+2t_2 Y+...)\\ &=&\det e^{ \sum_1^{\iy} t_i Y^i}
\prod^n_{i=1}(t_1+x_i(t_2,t_3,...)),
\end{eqnarray*}
and $p_n(t,z)$, as announced in (7.1) and (7.3). Note that the $n$
roots $x_i(t_2,t_3,....)$ of the characteristic equation in $t_1$
are solutions in $(t_2,t_3,...)$ of the $n$-particle Calogero-Moser
system with initial configuration coordinates $(x_1,... ,x_n,
\xi_1,...,\xi_n)$; see T. Shiota's paper \cite{S}. Thus, a solution of the discrete KP system
corresponds to a flag of Calogero-Moser system generated by one
pair of semi-infinite matrices $X$ and $Y$, given by (7.2), for
arbitrary large $n$. \qed

\noindent\remark  Observe that for $t=0$, the parameters $x$ and $z$ in
$$p_n(0,z)=\det (zI-  Y)
\det\left(I-
\left(xI-   X \right)^{-1}\left(zI -  Y\right)^{-1} \right).
$$
are interchangeable (except for the trivial factor $\det (zI- Y)$).
This must be compared to the results in \cite{S} and \cite{SW}.

\vspace{0.7cm}

\section{\bf Discrete KdV-solutions, with upper - triangular $L^2$
} 

\medbreak

Letting all points $p_i$ in the soliton example converge to $p$,
all points $q_i$ converge to $-p$, and all $\lambda_i$ converge to
$1$, the weights $\rho_k(z)$ take on the form (8.2) below. For
future use, define the functions:
\begin{eqnarray*}
f_{\ell}&=&p^{\ell}\sinh\sum_{{\rm
odd}}t_ip^i\quad\quad\ell\mbox{\,\,even}\\
&=&p^{\ell}\cosh\sum_{{\rm odd}}t_ip^i\quad\quad\ell\mbox{\,\,odd}
\end{eqnarray*}
and
\begin{eqnarray}
g_{\ell}&=&p^{\ell}\left(z\sinh\sum_{{\rm odd}}t_ip^i-p
\cosh\sum_{{\rm odd}}t_ip^i\right)\quad
\quad\ell\mbox{\,\,even}\nonumber\\
&=&p^{\ell}\left(z\cosh\sum_{{\rm odd}}t_ip^i-p\sinh\sum_{{\rm
odd}}t_ip^i\right)\quad\quad\ell\mbox{\,\,odd}.
\end{eqnarray}

\begin{theorem} The family of weights,
\be
\rho_k(z)=(-1)^k\dt^{(k)}(z-p)-\dt^{(k)}(z+p),~~\mbox{for}~0\leq k
\leq n-1,
\ee
leads to discrete KdV solutions, with KdV
$\tau$-functions\footnote{$W[...]$ denotes a Wronskian with respect
to the parameter $p$.}
\be
\tau_n(t)=2^n e^{n\displaystyle{\sum_{{\rm
even}}}t_ip^i}W[f_0,f_1,...,f_{n-1}].
 \ee
 The matrix $L$ has the property that $L^2$ is
 \underline{upper-triangular}, with polynomial eigenvectors
 $L(t)p(t,z)=zp(t,z)$, given by
\be
p_n(t,z)=\frac{W[f_0,f_1,...,f_{n-1}]}{W[g_0,g_1,...,g_{n-1}]}
\ee
in terms of (8.1); i.e., the polynomials $p_n(t,z)$ satisfy
\underline{3-step relations} of the following nature:
 $$
 z^2 p_n(t,z)=\al_n p_n+\beta_n p_{n+1}+p_{n+2}.
$$
  Then
\begin{eqnarray*}
\WR_n&=&\left\{f=\sum_0^{\iy}a_iz^i\mbox{\,\,such
that\,\,}f^{(k)}(p)-(-1)^kf^{(k)}(-p)=0,\quad 0\leq k\leq
n-1\right\}\\ &=&\{1,z^2,z^4,...\}\oplus z(z^2-p^2)^n\{1,z^2,z^4,...\}
\end{eqnarray*}
and
\begin{eqnarray*}
\WR^*_n&=&\left\{\int\frac{\rho_k(u)du}{z-u}=\left(
\frac{\pl}{\pl p}\right)^k\left(
\frac{2p}{z^2-p^2}\right),k=0,...,n-1\right\}\oplus \HR_+\\
&=&\left\{\left(\frac{\pl}{\pl p}\right)^k
\left(\frac{1}{z^2-p^2}\right),k=0,...,n-1\right\}\oplus \HR_+.
\end{eqnarray*}
\end{theorem}

\proof Indeed, the form of the flags $\WR_n$ and $\WR^*_n$ follow from
the general formulae (3.8) and (3.9). Therefore
\begin{eqnarray*}
\tau_n(t)&=& \det (\mu_{ij})_{0\leq i,j \leq n-1}\\
&=&\det\left(\left(\frac{\pl}{\pl
p}\right)^k\oint_{z=\iy}z^{n-j-1}g(z)
\frac{2p}{z^2-p^2}dz\right)_{0\leq k,j\leq n-1},\mbox{using
(3.8)}\\ &=&\det\left(\left(\frac{\pl}{\pl p}\right)^k\oint
z^{n-j-1}g(z)
\left(\frac{1}{z-p}-\frac{1}{z+p}\right)dz\right)_{0\leq k,j\leq
n-1}\\
&=&\det\left(\left(\frac{\pl}{\pl p}\right)^k\left(p^{n-j-1}e^{\sum
t_ip^i}-(-p)^{n-j-1}e^{\sum t_i(-p)^i}\right)\right)_{0\leq k,j\leq
n-1}\\
&=&\det\left(\left(\frac{\pl}{\pl
p}\right)^kp^{n-j-1}e^{\displaystyle{\sum_{{\rm even}}}t_ip^i}
\left(e^{\displaystyle{\sum_{{\rm
odd}}}t_ip^i}+(-1)^{n-j}e^{-\displaystyle{\sum_{{\rm odd}}}t_ip^i}
\right)\right)_{0\leq k,j\leq n-1}\\
&=&2^n e^{n\displaystyle{\sum_{{\rm
even}}}t_ip^i}W[f_0,f_1,...,f_{n-1}],
\end{eqnarray*}
which is formula (8.3).

In order to express the wave vector, one needs to compute

\vspace{0.3cm}

\noindent $\displaystyle{z^n\tau_n(t-[z^{-1}])}$\hfill
\begin{eqnarray*}
&=&z^n \det\left(\left(\frac{\pl}{\pl
p}\right)^k\left(p^{n-j-1}e^{\sum
t_ip^i}(1-\frac{p}{z})-(-p)^{n-j-1}e^{\sum
t_i(-p)^i}(1+\frac{p}{z})\right)\right)_{0\leq k,j\leq n-1}\\
&=&\det\left(\left(\frac{\pl}{\pl
p}\right)^kp^{n-j-1}e^{\displaystyle{\sum_{{\rm even}}}t_ip^i}
\left(e^{\displaystyle{\sum_{{\rm
odd}}}t_ip^i}(z-p)+(-1)^{n-j}e^{-\displaystyle{\sum_{{\rm
odd}}}t_ip^i}(z+p)
\right)\right)_{0\leq k,j\leq n-1}\\
&=&2^n e^{n\displaystyle{\sum_{{\rm
even}}}t_ip^i}W[g_0,g_1,...,g_{n-1}],
\end{eqnarray*}
from which formula (8.4) follows.

Also notice that from the form of $\WR_n$, we have
$$
z^2 \WR_n \subset \WR_n~~\mbox{and thus}~~ z^2 \WR^t_n \subset
\WR^t_n.
$$
It shows that the $\tau_n(t)$'s are KdV $\tau$-functions. This
fact, combined with
$$
\WR^t_n=(\Span\{\rho^t_0,\rho^t_1,...,\rho^t_{n-1}\})^{\bot}=
\Span\{p_n(t,z),p_{n+1}(t,z),...\} \subset \HR_+,
$$
leads to the $3$-step relation:
$$ z^2 p_n(t,z)=\al_n p_n+\beta_n p_{n+1}+p_{n+2},
$$
establishing the upper-triangular nature of $L^2$.
\qed

\remark Letting $p \rightarrow 0 $ in $p^{-n(n+1)/2}\tau_n(t)$ leads
to the rational KdV solutions, i.e., the Schur polynomials with Young
diagrams of type $\nu=(n,n-1,...,1)$.

\end{document}